\documentclass[twocolumn]{aastex63}

\usepackage{amsmath}
\usepackage{hyperref}

\accepted{June 29, 2021}
\submitjournal{ApJ}

\shorttitle{Light Echo Radiative Transfer}
\shortauthors{Ding et al.}
\graphicspath{{./}{figures/}}

\begin{document}

\title{Radiative Transfer Modeling of An SN~1987A Light Echo---AT2019xis}

\correspondingauthor{Jiachen Ding}
\email{njudjc@tamu.edu}

\author[0000-0003-4928-6698]{Jiachen Ding}
\affiliation{Department of Atmospheric Sciences,Texas A\&M University, \\
College Station, TX 77843, USA}

\author{Lifan Wang}
\affiliation{Department of Physics and Astronomy, Texas A\&M University, \\
College Station, TX 77843, USA}


\author{Peter Brown}
\affiliation{Department of Physics and Astronomy, Texas A\&M University, \\
College Station, TX 77843, USA}

\author{Ping Yang}
\affiliation{Department of Atmospheric Sciences,Texas A\&M University, \\
College Station, TX 77843, USA}







\begin{abstract}

We use a Monte Carlo radiative transfer model (MCRTM) to simulate the UBVRI light curves, images and linear polarization of a light echo from supernova SN~1987A in the Large Magellanic Cloud (LMC) using various dust cloud shapes, sizes, and optical properties. We compare the theoretical simulations to the observations of AT2019xis, a light echo detected at a large angular distance (4.05$^{'}$) from SN~1987A. We estimate the size and optical thickness of the dust cloud based on the simulation results and the observations of Optical Gravitational Lensing Experiment (OGLE-IV) Transient Detection System (OTDS) I-band light curve. The mass of the dust cloud is calculated using the estimated size, optical thickness and extinction coefficient. If the dust cloud is assumed to correspond to a gas-to-dust ratio of 300, the total mass of the dust cloud is approximately 7.8-9.3 \(M_\odot\). Based on these theoretical models, we show that the morphological shapes of the light echoes in the wavelength range in or shorter than the U-band to be very different from those in the longer wavelength bands, and the difference carries important information on the early UV radiation of SN~1987A.

\end{abstract}



\section{INTRODUCTION} \label{sec:intro}
A light echo (LE) is the scattered star light by an interstellar or circumstellar dust cloud. The scattered photons of a light echo travel along different paths from the direct line of sight, so a light echo is always observed at a time after the source light arrives at the observer. Sometimes more than one light echo is observed at the same time. In this case, the scattered photons of each light echo propagate the same distance from the source to the observer but do not travel via the same path.

\begin{figure}[ht!]
\plotone{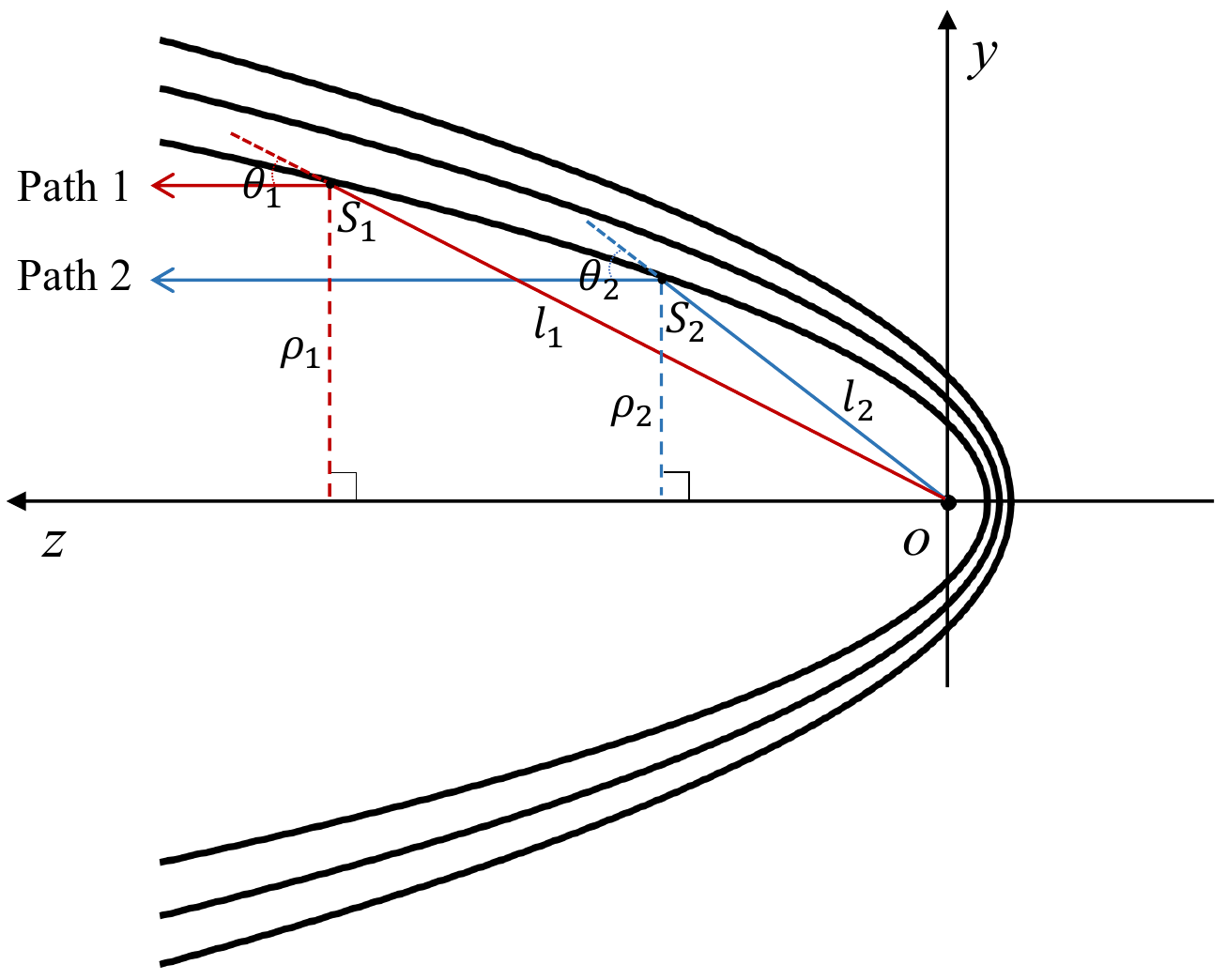}
\caption{Parabola illustration of light echo geometry in 2D. The light source is at origin $O$. The light source-observer line of sight is along the $z$ direction, and delay time is zero at any point on the $z$-axis. The three parabolas correspond to three nonzero delay times, with the inner parabola having a smaller delay. Two scattering points ($S_j, j=1,2$) and their light travel paths are plotted on the inner parabola. For scattering point $S_j$, $l_j$ is the distance from $O$ to $S_j$, ${\rho}_j$ is the orthogonal distance from $S_j$ to the light source’s line of sight on the $z$-axis, and ${\theta}_j$ is the scattering angle along that light travel path. Each point on a parabola has the same delay time.\label{fig:parabo}}
\end{figure}

The light echo generation mechanism and geometry can be described by an ellipsoid model if only single scattering is considered. The light source and observer are at two foci of a prolate spheroid. The distance from the light source to any point on the spheroidal surface plus the distance from the point to the observer is a constant, which corresponds to a single delay time. In practical astronomical observation, the observer can be assumed to be infinitely far from the light echo source. The ellipsoid model can thus be simplified to a paraboloid model \citep{Couderc1939}. Figure \ref{fig:parabo} illustrates the paraboloid model in 2D, which is a parabola in the $yz$-plane. The intersection of any paraboloid with the $xy$-plane, corresponding to a telescope field of view, is a circle centered on the $z$-axis at the source coordinate.

For any parabola in Figure \ref{fig:parabo}, the delay time is $t_{d}$, the distance from the light source to a scattering dust cloud is $l$, the distance between the scattering dust cloud and the light source’s line of sight is $\rho$, and the dust cloud scattering angle is $\theta$. Based on the parabola model, we have the following relations for any scattering object:
\begin{subequations} \label{eq:parabo}
\begin{equation}
l=\frac{1}{2}\left(\frac{{\rho}^2}{ct_d}+ct_d\right),
\end{equation}
\begin{equation}
\cos{\theta}=\frac{{\rho}^2-(ct_d)^2}{{\rho}^2+(ct_d)^2},
\end{equation}
\end{subequations}

\noindent
where $c$ is the speed of light. $l$ and $\theta$ can be estimated based on Equation \eqref{eq:parabo} if $\rho$ and $t_d$ are known. $\rho$ and $t_d$ can be accurately measured in light echo observation, provided that the light source is identified. More detailed descriptions of the light echo paraboloid model have been reported in previous studies (e.g., \citealt{Chevalier1986,Sparks1994,Sugerman2003,Tylenda2004,Patat2005}). Single scattering usually dominates the observed light echo signal so the paraboloid model is still useful to analyze light echo observations with some multiple scattering contribution. In Figure \ref{fig:parabo}, the three parabolas correspond to three delay times. Although Paths 1 and 2 are different, the scattering points $S_1$ and $S_2$ are on the same parabola, and the path lengths to the observer are the same, so the scattered light traveling via these paths can be observed at the same time.

While any star can produce a light echo, a substantial amount of information can be obtained if the source light comes from transients such as supernovae (SNe) and novae, or other variable objects such as Miras and Cepheids \citep{Sugerman2003}. A light echo contains information such as spectra of the source light that illuminated the dust cloud, so light echoes can be used to trace the characteristics of historical SNe (e.g., \citealt{Rest2005,Dwek2008,Rest2008,Rest2012}), which have exploded many years ago. On the other hand, a light echo also carries fruitful information about the dust cloud \citep{Rest2012a} including optical properties, grain size and composition.

Understanding the properties of interstellar dust is important from various aspects of astronomical research. This is because astronomical observations are usually affected by interstellar dust extinction along the path to the earth. To obtain the absolute brightness of a star such as a supernova, we need to correct for extinction by interstellar dust. For example, Type Ia SNe can be used as extragalactic distance indicators (e.g., \citealt{Riess1996,Perlmutter1998}) in cosmology studies. The uncertainties in distance estimation using Type Ia SNe can be reduced if dust extinction is accurately accounted for. Interstellar dust also plays critical roles in star formation and galaxy evolution.

Interstellar dust extinction can be directly measured along the line of sight (e.g., \citealt{Clayton1985,Fitzpatrick1986,Cardelli1988}). In observations along the line-of-sight, the scattered light cannot be fully separated from the directly transmitted light and the dust property information is only limited to a single sightline. In contrast, light echoes may contain scattered light alone along various photon travel paths. The two-dimensional image and even three-dimensional structure of a dust cloud can be inferred by light echo observations. Previous studies (e.g., Nova Persei 1901 LEs by \citealt{Couderc1939}; SN~1987A LEs by \citealt{Suntzeff1988,Crotts1988,Xu1995}; SN 2006X LEs by \citealt{Wang2008a}; SN 2007af LEs by \citealt{Drozdov2015}; SN 2014J LEs by \citealt{Crotts2015,Yang2017} to list just a few) successfully utilize light echoes to infer interstellar and circumstellar dust properties.

Considering interstellar dust concentration and optical properties, the brightness of a light echo is approximately 10 magnitudes fainter than the source light that illuminates the dust cloud \citep{Patat2005}. Because supernovae can be highly luminous at their peak brightness and their explosion is like a flash, the light echoes induced by supernovae are easier to observe and are used to study interstellar/circumstellar dust properties. The Type II SN~1987A \citep{Arnett1989,McCray2016} in the Large Magellanic Cloud (LMC) is the closest to the earth supernova that people have observed since Kepler’s Supernova. Many SN~1987A light echoes have been captured and utilized to infer the characteristics of SN~1987A and its nearby dust.

Light echoes scattered by interstellar dust about 100 to 400 pc in front of SN~1987A are first observed and identified using imaging and spectroscopy \citep{Crotts1988,Suntzeff1988,Gouiffes1988,Couch1990} as predicted by \cite{Schaefer1987,Chevalier1988}. At distances several arcseconds to SN~1987A, \cite{Crotts1989} discovered diffuse radiations which are well resolved to be nebular rings by images taken at the ESO NTT \citep{Wampler:87A:1990ApJ...362L..13W,Wang:Napoleon:1992A&A...262L...9W}. \cite{Wampler:87A:1990ApJ...362L..13W} show also evolving diffuse light immediately outside the nebular rings which is known as the Napoleon's Hat nebula \citep{Wang:Napoleon:1992A&A...262L...9W}. While the rings are modelled in terms of the interaction between the low velocity wind presumably from the progenitor during the red supergaint (RSG) phase with a subsequent energetic fast wind during the blue supergiant (BSG) phase \citep{Wang:Mazzali:1992Natur.355...58W}, the Napoleon's Hat nebula is found to be produced by dusty materials lost during the RSG, and the bow-shock-like morphology is caused by a  differential motion between the progenitor star and the interstellar matter \citep{Wang:Napoleon:1992A&A...262L...9W,Wang:Napolean1993MNRAS.261..391W}.  \cite{Xu1995} mapped the 3D interstellar dust structure in front of SN~1987A using more than 5 years of light echo images. With the decrease of SN~1987A brightness, more light echoes scattered by circumstellar dust were also observed and resolved  \citep{Wampler:87A:1990ApJ...362L..13W,Crotts1989,Bond1990,Wang:Napoleon:1992A&A...262L...9W}. SN~1987A light echo images are used to probe the three-ring circumstellar dust structure around SN~1987A (e.g., \citealt{Wang:Napolean1993MNRAS.261..391W,Sugerman2005,Sugerman2005a}). Model calculations suggest that circumstellar dust scattering or light echo may significantly contribute to the observed SN~1987A polarization \citep{Wang1996}.

A recently observed transient, AT2019xis in the Large Magellanic Cloud (LMC) has been identified as a light echo of SN~1987A \citep{Taubenberger2019} through I-band light curve and spectrum comparisons. The angular distance between SN~1987A and AT2019xis is about 4.05$'$ and the geometric distance is about 200 pc or $\approx$650 ly \citep{Taubenberger2019}. Represented schematically by Path 1 in Figure \ref{fig:parabo}, $l_1$ is the distance from SN~1987A to AT2019xis, and the Path 1 and $y$-axis arrows intersect at the earth at a 4.05$'$ angle. This angular distance is far larger than previously observed Napolean's Hat \citep{Wang:Napoleon:1992A&A...262L...9W,Wang:Napolean1993MNRAS.261..391W}, so the impact of directly transmitted source light can be completely excluded and the dust scattering angle is large ($\approx$20$^{\circ}$, angle ${\theta}_1$ in Figure \ref{fig:parabo}). Because the scattering angle is large, the strong forward scattering by the dust cloud hardly contributes to the light echo signal and the light echo brightness must be far weaker than the source light. Thus, the supernova needs to be very nearby to make such a light echo observable. AT2019xis is a unique light echo to study interstellar dust optical properties at a large scattering angle.

\begin{figure}[ht!]
\plotone{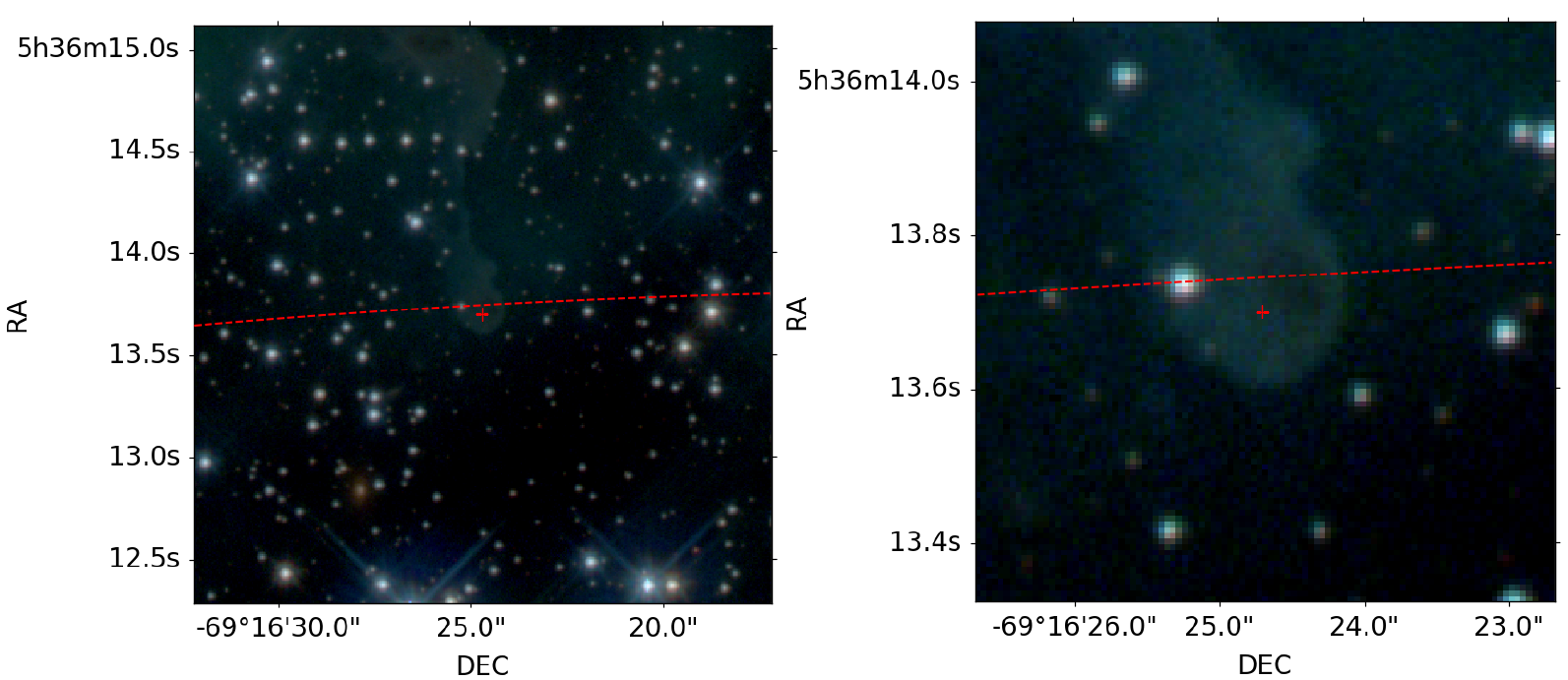}
\caption{RGB images from 2014 showing the region where AT2019xis was observed. The RGB images are plotted using HST WFC3 data. Astronomical North (increasing DEC) is at the right of each image and astronomical East (increasing RA) is at the top. The red cross denotes the coordinate of AT2019xis (RA 05:36:13.700 DEC -69:16:24.70). The red dashed line across the middle of each panel is the intersection between the field of view and the SN~1987A light echo paraboloid at the distance and declination of AT2019xis. \textbf{Left image}: measures 15$''$ $\times$ 15$''$; \textbf{Right image}: measures 4$''$ $\times$ 4$''$.\label{fig:hstimg}}
\end{figure}

Figure \ref{fig:hstimg} shows RGB images from August 2014 where AT2019xis was observed in 2019. The images are composed from archival Hubble Space Telescope (HST) Wide Field Camera 3 (WFC3) UVIS channel observations  (\# 13401, PI: Claes Fransson) obtained from Mikulski Archive for Space Telescopes (MAST; no comparable images showing the AT2019xis light echo are available). A detailed list of the images is in the appendix \ref{apdx:HSTdata}. The red (R), green (G), and blue (B) channels use HST data from F814W, F606W and F475W filters respectively. The medians of the pixel values in images of an individual filter are used to construct the image in a color channel. Because at the time the HST images were taken, the SN~1987A emission has not yet arrived at the AT2019xis location, the signal in Figure \ref{fig:hstimg} is emission by the dust cloud. Around the location of AT2019xis, the dust cloud has a disk-like shape with a nonconcentric hole. The left plot of Figure \ref{fig:hstimg} shows that the disk-like structure is part of a larger dust cloud. The light echo was from a reflective patch with dimension $\approx$0.6 ly \citep[Figure \ref{fig:hstimg};][]{Taubenberger2019}, or about half of the width of the visible dust cloud around AT2019xis, so other light echoes may be observed in the future from the remaining part of the dust cloud.

 The AT2019xis dust cloud has a similar distance to the SN~1987A along the line of sight (about 170 pc) as the previously discovered circular echoes (e.g., \citealt{Crotts1988,Suntzeff1988,Gouiffes1988,Couch1990}). However, the AT2019xis dust cloud has a much larger angular distance (approximately 4.05$'$) from the supernova. If we assume a light echo by a dust cloud that is at the same distance along the line of sight to the SN~1987A as the AT2019xis dust cloud is observed one year after the SN~1987A explosion, the distance between the AT2019xis dust cloud and the assumed dust cloud is about 50 pc. Therefore, the AT2019xis dust cloud may be part of a superbubble that is in front of SN~1987A.

Models of the light echo brightness and time evolution are always needed to infer meaningful information from the light echo observations. Single-scattering approximation can be used in modeling light echoes (e.g., \citealt{Chevalier1986,Schaefer1987,Wang1996,Sugerman2003,Patat2005}) if the dust cloud is optically thin. Under the single-scattering approximation, the light echo brightness is expressed as a product of source light flux, dust scattering phase function and scattering efficiency, and dust concentration. The ellipsoid or paraboloid model is used to analyze light echo time evolution. If the dust cloud is optically thick and dense such as is the dust cloud of AT2019xis, multiple scattering contributions must be included. Monte Carlo techniques have been used to simulate the radiative transfer process of light echoes (e.g., \citealt{Witt1977,Chevalier1986,Wang:2005ApJ...635L..33W,Patat2005,Patat2006}), which can take into account multiple scattering effects.

In this study, we use a Monte Carlo radiative transfer model (MCRTM) to simulate the observation of an interstellar dust cloud with similar observing geometry to AT2019xis, and show how the simulated observations vary with different dust optical properties, dust cloud sizes and shapes. We also estimate the dust cloud properties using our simulation results and I-band light curve observations from the Optical Gravitational Lensing Experiment (OGLE-IV) Transient Detection System (OTDS; \citealt{Kozlowski2013,Wyrzykowski2014}). In the following sections, we introduce the methodology and observations in section \ref{sec:method}, present the simulation and estimation results in section \ref{sec:results}, and summarize the study in section \ref{sec:summary}.

\section{METHODOLOGY} \label{sec:method}

\subsection{Radiative Transfer Simulation} \label{subsec:MCRTM}

The MCRTM is used to simulate the observation of the SN~1987A light echo by the dust cloud at AT2019xis location and outputs the light echo’s Stokes vector. A detailed description of the MCRTM is in appendix \ref{apdx:MCRTM}.

\begin{figure}[ht!]
\plotone{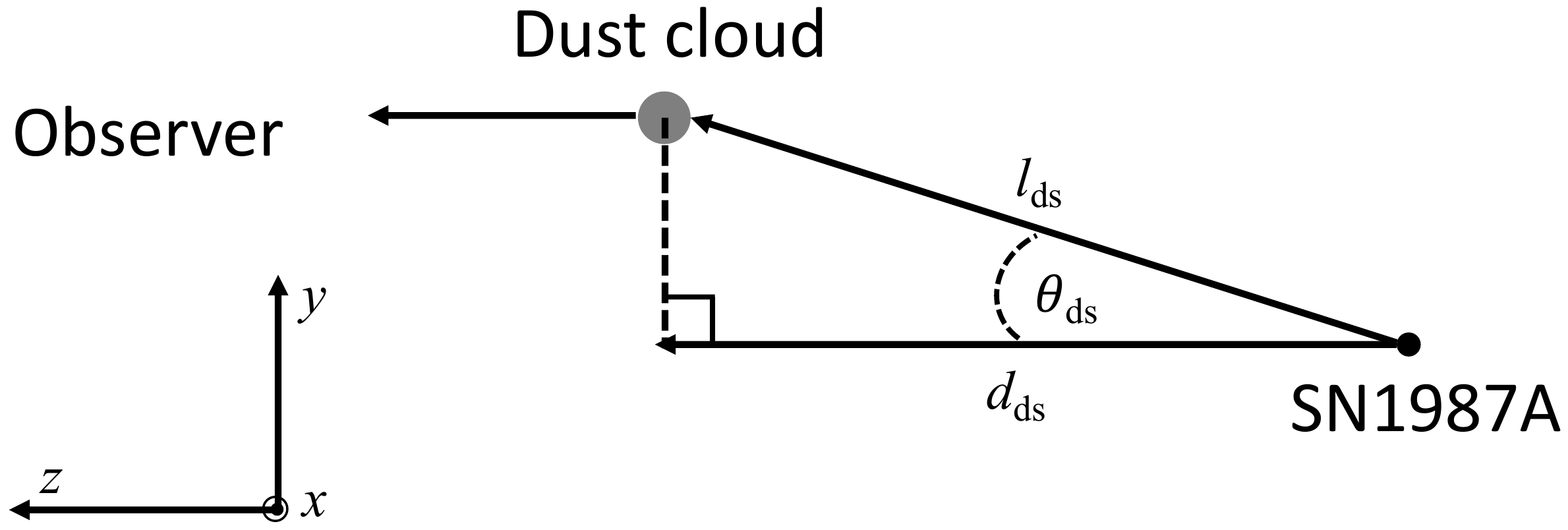}
\caption{Modeled observational geometry. SN~1987A is at the origin. $l_\text{ds}$ is the distance between the dust cloud center and SN~1987A. $d_\text{ds}$ is the distance between the dust cloud and SN~1987A along the line of sight or $z$- direction. ${\theta}_\text{ds}$ is the viewing angle looking from SN~1987A to the dust cloud and is also the single-scattering angle.\label{fig:obsg}}
\end{figure}

The modeled observational geometry is illustrated in Figure \ref{fig:obsg}. The dust cloud and SN~1987A are in the $yz$-plane. The polarization reference plane of the observed Stokes vector is the $xz$-plane. The $xy$-plane is divided into a variety of square bins to simulate the light echo image. At each scattering position, if the $x$ and $y$ coordinates are within a predefined square bin, the contribution of the photon is counted to the light echo at the center of this square bin.

\subsection{Interstellar Dust Optical Properties} \label{subsec:dust}

In the radiative transfer simulation involving a dust cloud, the interstellar dust optical properties including extinction cross section ($C_\text{ext}$), SSA and scattering phase matrix are needed. In this study, we adopt the optical properties of the dust grain model developed by \citet{Weingartner2001}, \citet{Li2001} and \citet{Draine2007} (Hereafter, WD01 model).

The WD01 dust grain model assumes the dust grain is a mixture of carbonaceous and silicate particles and has a grain-size distribution with more than ten adjustable parameters. The grain size ranges from 3.5 \r{A} - 10 ${\mu}m$. The dust particle shape is assumed to be a sphere. Mie theory is utilized to obtain the single-scattering properties of the dust particle \citep{Weingartner2001}. The dust grain model can reproduce observed interstellar dust extinction and emission \citep{Weingartner2001,Li2001} by adjusting the grain-size distribution parameters. In this study, we consider dust models in the Large Magellanic Cloud (LMC), Small Magellanic Cloud (SMC), and Milky Way (MW) including six models, namely “LMC avg”, “LMC 2”, “SMC bar”, and three MW dust models with different total-to-selective extinction ratios ($R_\text{V}$).

\begin{figure}[ht!]
\plotone{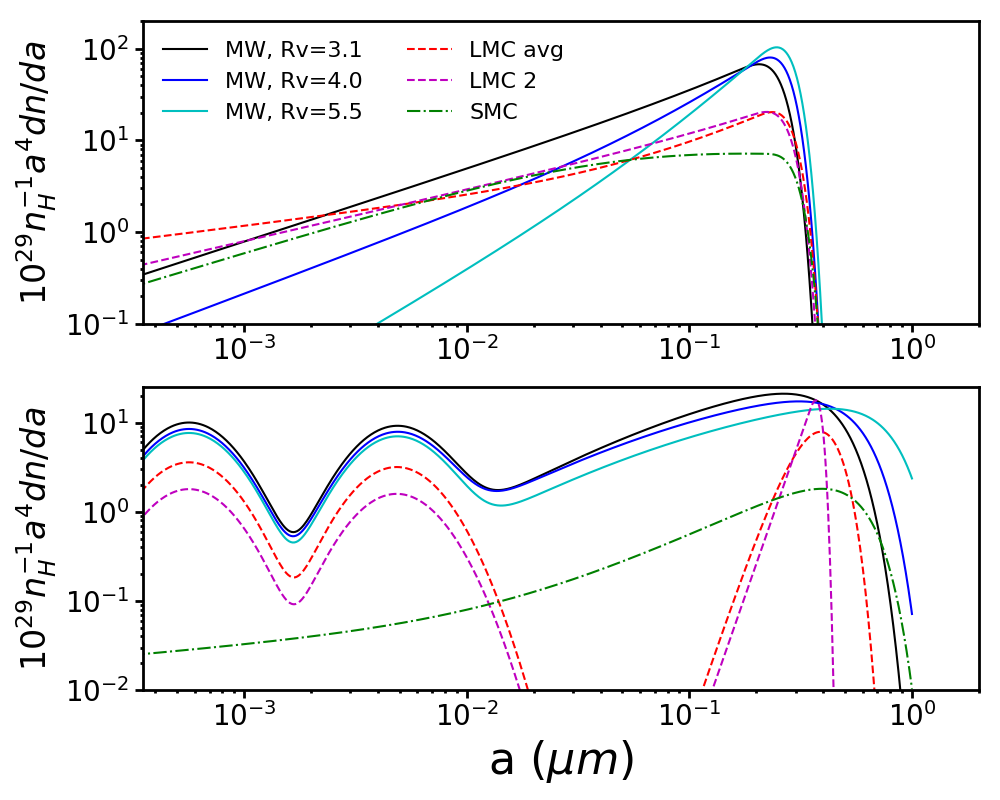}
\caption{WD01 dust model size distributions. In the plot, the size distributions are weighted by $a^4$ for visualization purposes, where $a$ is the grain size. \textbf{Upper plot}: silicate particle; \textbf{Lower plot}: carbonaceous particle.\label{fig:sized}}
\end{figure}

Each dust grain model in \citet{Weingartner2001} has multiple sets of grain-size distribution parameters. As suggested by \citet{Weingartner2001},  the grain-size distribution parameter sets with relatively large C abundance per H nucleus ($b_c$) are favored, which can better fit the extinction curve 2175 Angstrom hump. We use the “LMC avg”, “LMC 2”, “SMC bar”, “MW, $R_\text{V}$=3.1”, “MW, $R_\text{V}$=4.0” and “MW, $R_\text{V}$=5.5” models with $b_c$ values $2.0\times10^{-5}$, $1.0\times10^{-5}$, $0$, $5.58\times10^{-5}$, $4.72\times10^{-5}$ and $4.26\times10^{-5}$ respectively, in which the MW models’ $b_c$ values are adjusted (\citealt{Draine2003a,Draine2003b}) from the original \citet{Weingartner2001} $b_c$ values. The size distributions of the six dust models are shown in Figure \ref{fig:sized}. For MW dust, the models with larger $R_\text{V}$ have a larger portion of large-sized particles.

\begin{figure}[ht!]
\plotone{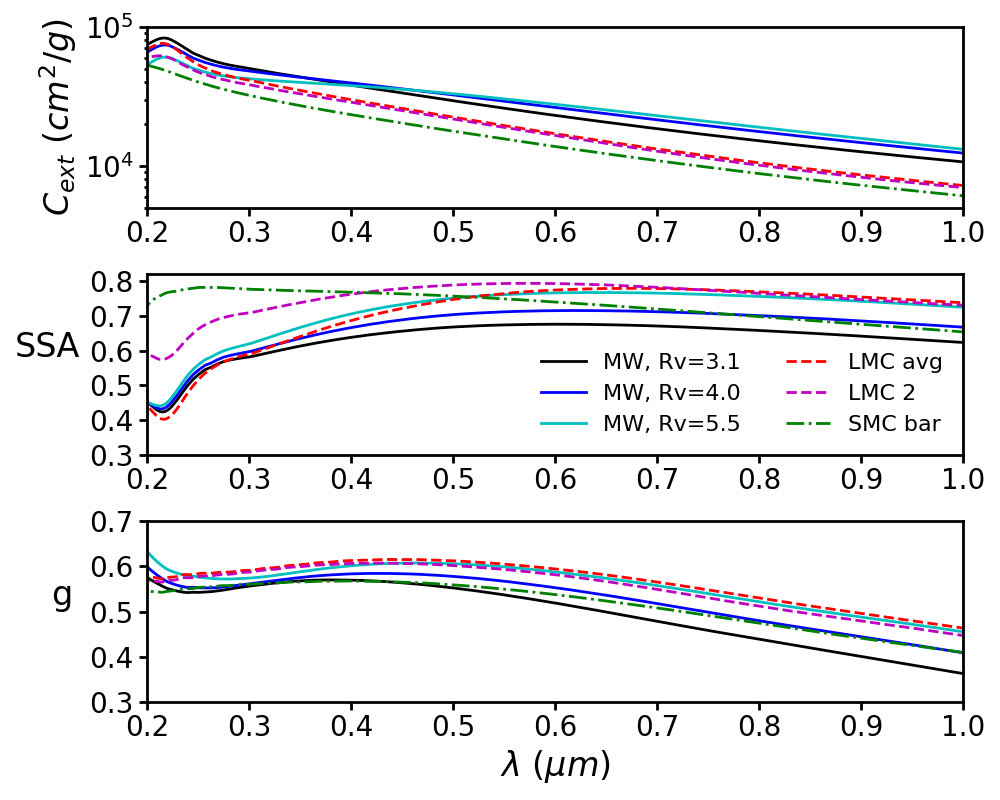}
\caption{WD01 interstellar dust optical properties.\label{fig:dustp}}
\end{figure}

The spectral optical properties including $C_\text{ext}$, SSA and asymmetry factor ($g$) from ultraviolet (UV) to near infrared (NIR) are shown in Figure \ref{fig:dustp}. From red to NIR wavelengths, “MW, $R_\text{V}$=5.5”, “LMC avg”, “LMC 2” have very similar SSA and $g$, which are larger than other models.

In the computation, we specify the optical thickness or optical depth ($\tau$) at one wavelength ${\lambda}_0$, and use the following equation to obtain the optical thickness at other wavelengths:
\begin{equation} \label{eq:taulambda}
\tau(\lambda)=\frac{C_{\text{ext,}\lambda}}{C_{\text{ext,}{\lambda}_0}}\tau({\lambda}_0).
\end{equation}

Because the WD01 dust models do not provide full scattering phase matrix data, we use the Henyey-Greenstein (H$\text{-}$G) phase function \citep{Henyey1941} with the WD01 asymmetry factor data. A Rayleigh scattering phase matrix is utilized to simulate polarization, by keeping $F_{ij}/F_{11}$ ($i\text{, }j=1\text{, }2\text{, }3\text{, }4$) unchanged and replacing $F_{11}$ by the H$\text{-}$G phase function, where $F_{ij}$ are scattering phase matrix elements.

\subsection{Comparison with Light Curve Observation} \label{subsec:otds}

We compare MCRTM simulation with the OGLE-IV OTDS I-band light curve observation of AT2019xis. The AT2019xis I-band light curve data are downloaded from the OGLE-IV OTDS website (\url{http://ogle.astrouw.edu.pl/ogle4/transients}). The OGLE-IV is the fourth phase of the OGLE sky survey project \citep{Udalski2015}. The OGLE-IV OTDS is one of the OGLE-IV services aiming at observing supernovae and other transients in the Magellanic Cloud Systems \citep{Kozlowski2013,Wyrzykowski2014}. The OGLE-IV observation is carried out by a 1.3 m Warsaw telescope at the Las Campanas Observatory, Chile. This telescope is equipped with a 32 CCDs camera \citep{Udalski2015}. The OGLE-IV uses two filters of Johnson-Cousin I and V bands. The observation cadence for LMC is around 5 days. The detectable I-band magnitude can be as faint as 22. Data reduction is conducted near real time at the telescope site and then the reduced data are fed into the OGLE photometric pipeline \citep{Wyrzykowski2014}. The photometry of ongoing transients is available at the above-mentioned website, where the I-band magnitude is roughly calibrated within 0.2 \citep{Wyrzykowski2014}.

The OGLE-IV OTDS I-band has center wavelength around 0.8 ${\mu}m$. The simulations are performed from 0.2 to 1.0 ${\mu}m$ at wavelengths in WD01 data. The band-averaged Stokes vector is computed by
\begin{equation} \label{eq:stkavg}
\bar{\bf{I}}=\frac{{\int}T(\lambda)\,\bf{I}(\lambda)\,\emph{d}\lambda}{{\int}T(\lambda)\,d\lambda},
\end{equation}

\noindent
where $T$ is filter transmission and $\bf{I}$ is the Stokes vector.

\begin{figure}[ht!]
\plotone{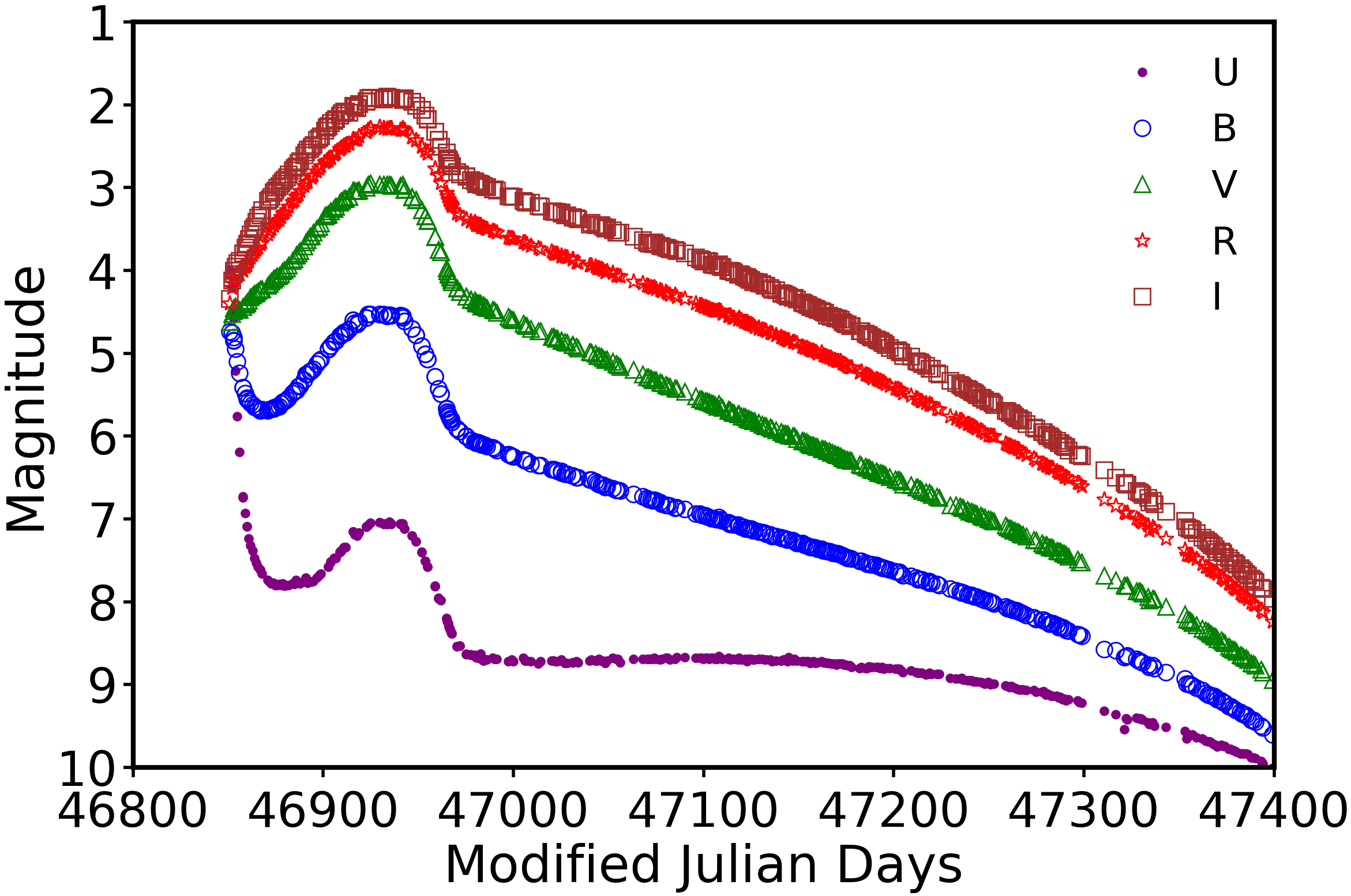}
\caption{SN~1987A UBVRI light curves.\label{fig:SN87A}}
\end{figure}

In our MCRTM computations, the photons are emitted at one single time so the results are equivalent to an impulse response with a Dirac Delta function as the input supernova light curve. To compare with the observed AT2019xis light echo light curve, we need to consider SN~1987A light curves as the input. The simulated SN~1987A light echo Stokes vector is a convolution between the impulse response Stokes vector and the SN~1987A light curve, expressed as
\begin{equation} \label{eq:stkle}
\bar{\bf{I}}_{LE}(t)={\int}\bar{\bf{I}}(t-t')\,\bar{F}_{\text{SN}}(t')\,dt',
\end{equation}

\noindent
where $\bar{\bf{I}}_{LE}$ is the simulated light echo Stokes vector and $\bar{F}_{\text{SN}}$ is the supernova flux at a specific band. The SN~1987A light curve data are obtained from the first 500 days' photometric observation of SN~1987A at the Sutherland field station of South African Astronomical Observatory \citep{Menzies1987,Catchpole1987,Catchpole1988,Catchpole1989,Whitelock1988}, and downloaded from Open Supernova Catalogue  (OSC; \citealt{Guillochon2017}\footnote{https://sne.space}). The observed UBVRI
light curves are shown in Figure \ref{fig:SN87A}.

The parameters $l_{\text{ds}}$, $d_{\text{ds}}$ and $\theta_{\text{ds}}$ in Figure \ref{fig:obsg} can be determined if we know the light delay time $t_d$ and the angular distance between AT2019xis and SN~1987A ${\alpha}_\text{ds}$, both of which are measured at the earth. $l_{\text{ds}}$, $d_{\text{ds}}$ and $\theta_{\text{ds}}$ have the following geometric relation
\begin{subequations} \label{eq:geom}
\begin{equation}
\left(l_\text{SN}-d_\text{ds}\right)\tan{\alpha_\text{ds}}=l_{\text{ds}}\sin{\theta_{\text{ds}}},
\end{equation}
\begin{equation}
ct_d=l_{\text{ds}}-d_{\text{ds}},
\end{equation}
\end{subequations}

\noindent
where $l_\text{SN}$ is the distance between the earth and SN~1987A and ${\alpha}_\text{ds}$ ($'$) is determined by the coordinates of SN~1987A (RA 05:36:13.700 DEC $-$69:16:24.70) and AT2019xis (RA 05:35:27.989 DEC $-$69:16:11.50). The delay time $t_d$ is about 32.65 years according to observed dates of SN~1987A and AT2019xis. We use cross correlation between observed and simulated light echo light curves to obtain a more accurate $t_d$. $\theta_{\text{ds}}$ and $l_{\text{ds}}$ are thus determined to be 18.9$^{\circ}$ and 186.3 pc, and $t_d$ is computed to be 32.88 years, with the difference from 32.65 years probably resulting from imprecision in all measurements and computations.

We also estimate the dust cloud geometric size and optical thickness from the observed AT2019xis light curve in the next section. We use the following cost function:
\begin{equation} \label{eq:costf}
J=\left[L_{\text{obs}}-L_{\text{sim}}(\tau,r)\right]^T{\bf{S}}^{-1}_{\epsilon}\left[L_{\text{obs}}-L_{\text{sim}}(\tau,r)\right],
\end{equation}

\noindent
where $r$ is the characteristic size of the dust clump, $L$ is the light curve magnitude and 
$\bf{S}_{\epsilon}$ is the covariance matrix of observational errors. Here $\bf{S}_{\epsilon}$ is assumed to be a diagonal matrix with measurement uncertainty as the diagonal elements. We use the Levenberg-Marquardt method \citep{Rodgers2000} to minimize the cost function $J$ and obtain the best-fit $\tau$ and $r$.

\section{RESULTS} \label{sec:results}

\begin{figure}[ht!]
\plotone{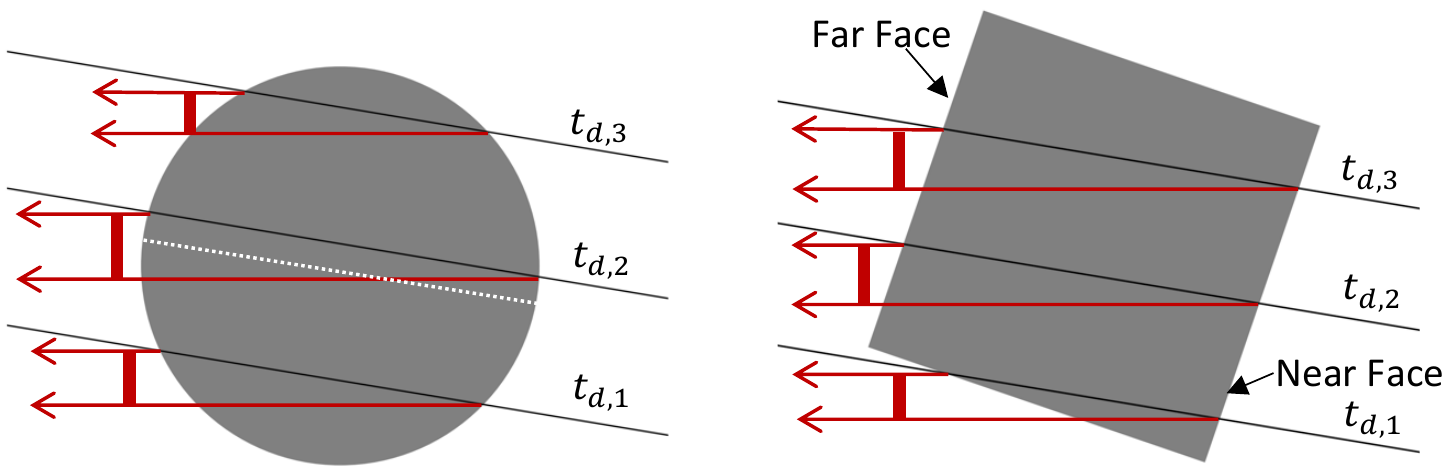}
\caption{Two-dimensional views ($yz$-plane, consistent with Figure \ref{fig:obsg}) of the dust cloud shapes used in the MCRTM simulation. The black curves are light echo parabolas corresponding to three delay times $t_{d,1}$, $t_{d,2}$ and $t_{d,3}$. $t_{d,1}<t_{d,2}<t_{d,3}$. The red arrows are along the line of sight toward the observer. The red rectangles denote the areas where the dust cloud is illuminated at a single delay time. $\textbf{Left plot}$: A spherical dust cloud. The white dashed line denotes a great circle of the sphere. $\textbf{Right plot}$: A cube-shaped dust cloud. The annotated Near Face and Far Face are two faces of the cube perpendicular to the $yz$-plane. \label{fig:view2D}}
\end{figure}

The simulation assumes two cloud geometries, namely a spherical cloud and cube-shaped cloud. Apparently, neither shape is a realistic shape for the dust cloud as shown in Figure \ref{fig:hstimg}. but they are selected to study the sensitivity of a light echo to the dust cloud shape. For the cube shape, one face of the cube is perpendicular to the straight line formed by SN~1987A and the geometric center of the cube. A 2D view of the two dust cloud shapes overlapping with the parabola model is shown in Figure \ref{fig:view2D}. If the light source is an impulse, the singly scattered light echo can be predicted using the paraboloid model.

As shown in Figure \ref{fig:view2D}, the simulated impulse response of singly scattered light will be observed as a bright strip with certain width at a single delay time. The shape of the observed bright strip is determined by the dust cloud shape. For a spherical shape, the length and width of the strip keep increasing until the light echo paraboloid intersects the great circle of the sphere. Then, the length and width of the strip start to decrease. For a cube, because of the orientation of the cube, the length of the bright strip will almost be a constant. The width of the bright strip increases until the light echo paraboloid intersects both the Near Face and Far Face annotated in Figure \ref{fig:view2D}. Then, the width of the strip is a constant until the light echo paraboloid does not intersect the Near Face, and the width then starts to decrease. The variation of the light echo shape will be reflected on the light curve shape. When the light curve of the source supernova is included and multiple scattering is considered, the light echo variation will be more complicated.

\subsection{Light Curve Simulations} \label{subsec:lcs}

\begin{figure}[ht!]
\plotone{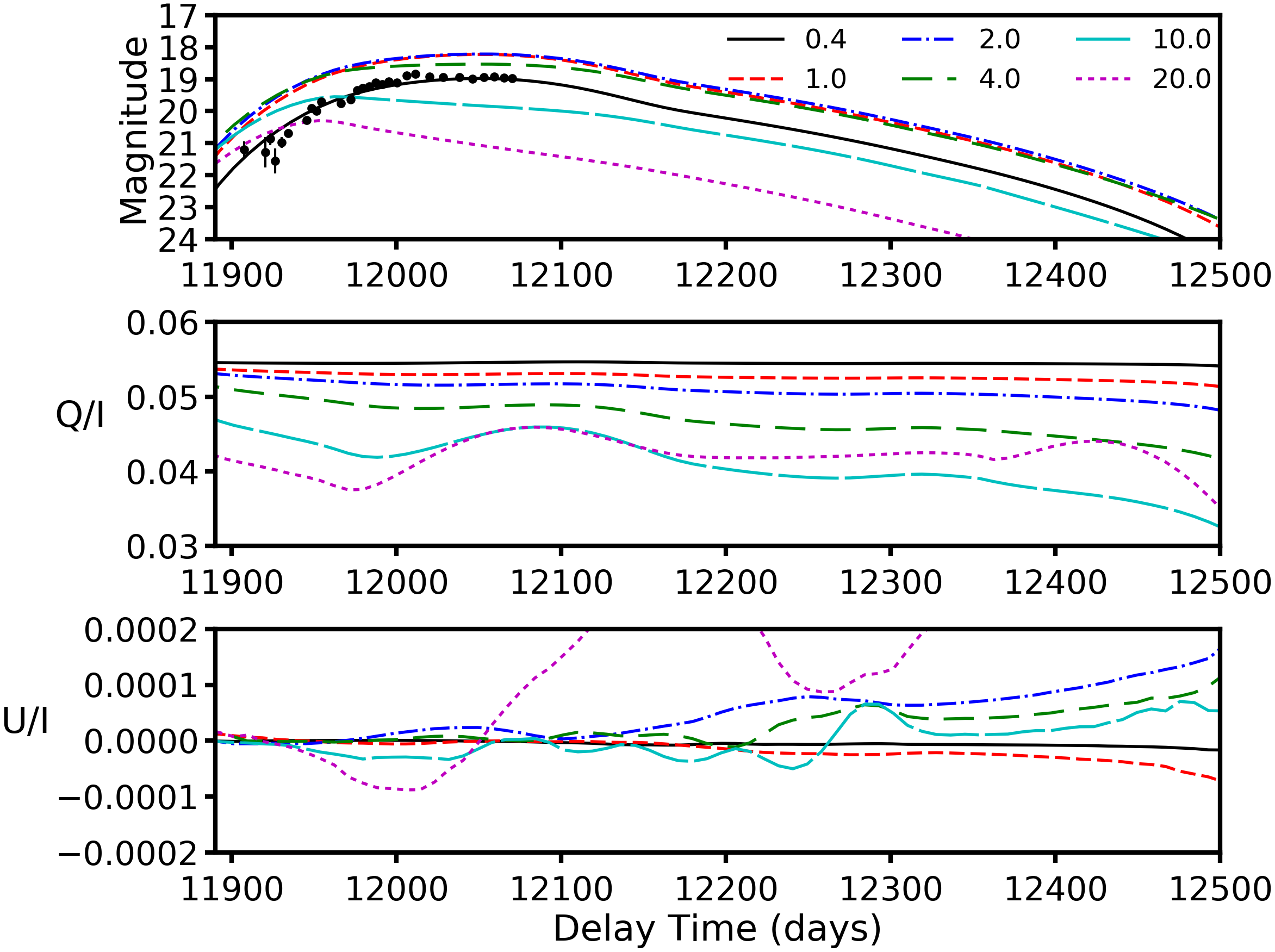}
\caption{Simulated SN~1987A I-band light echo light curves and associated $Q$ and $U$ components of Stokes vector with varied diameter optical thickness at 0.8 ${\mu}m$ wavelength. The dust cloud is assumed to be a sphere having diameter 1.8 light years (ly). The “LMC avg” dust model is used in the simulation. The black dots are AT2019xis I-band observations by OGLE-IV OTDS. \label{fig:lc1}}
\end{figure}

\begin{figure}[ht!]
\plotone{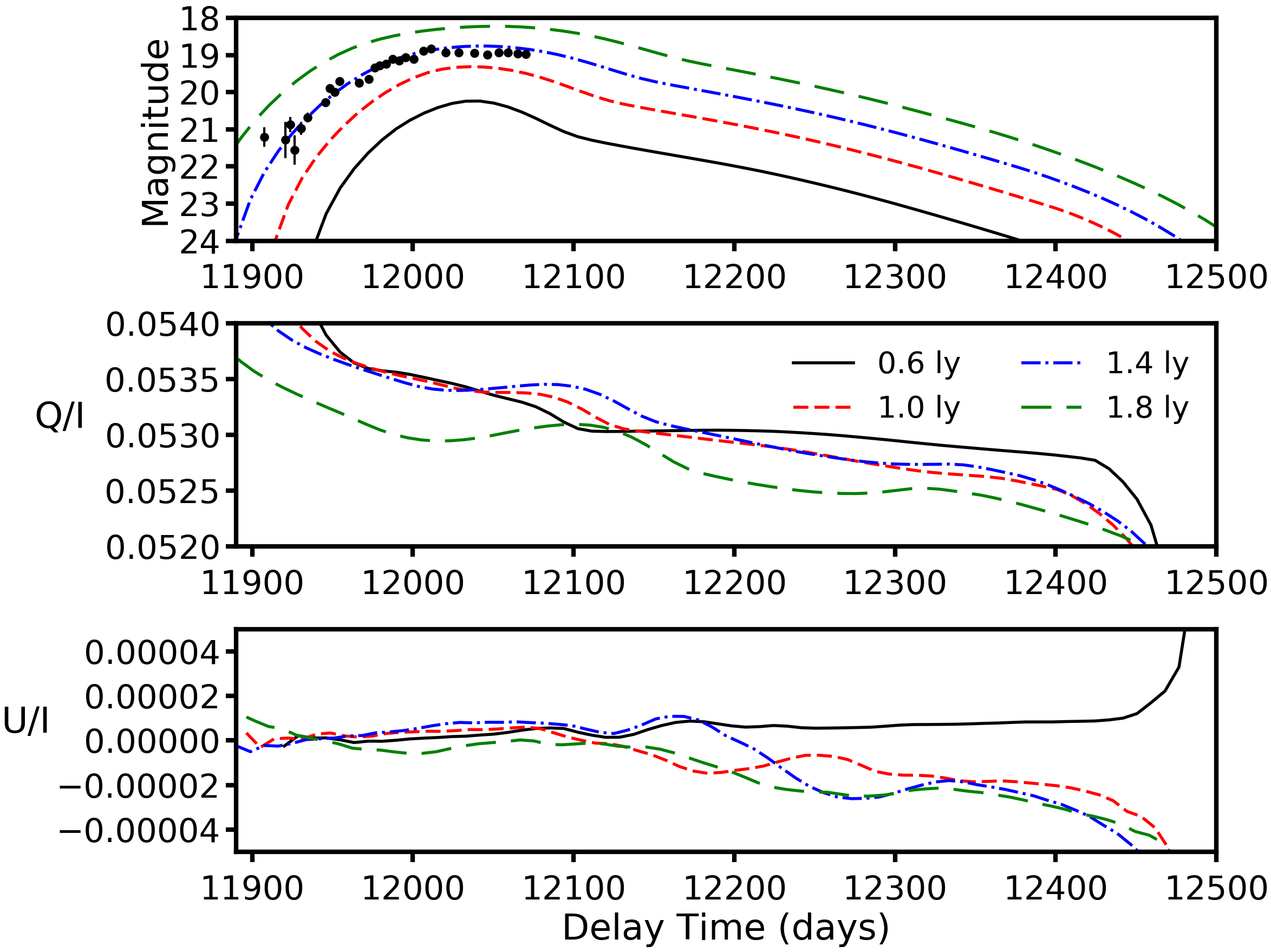}
\caption{The same as Figure \ref{fig:lc1}, but with varied dust cloud sphere diameters. The diameter optical thickness at 0.8 ${\mu}m$ wavelength is 1.0. \label{fig:lc2}}
\end{figure}

\begin{figure}[ht!]
\plotone{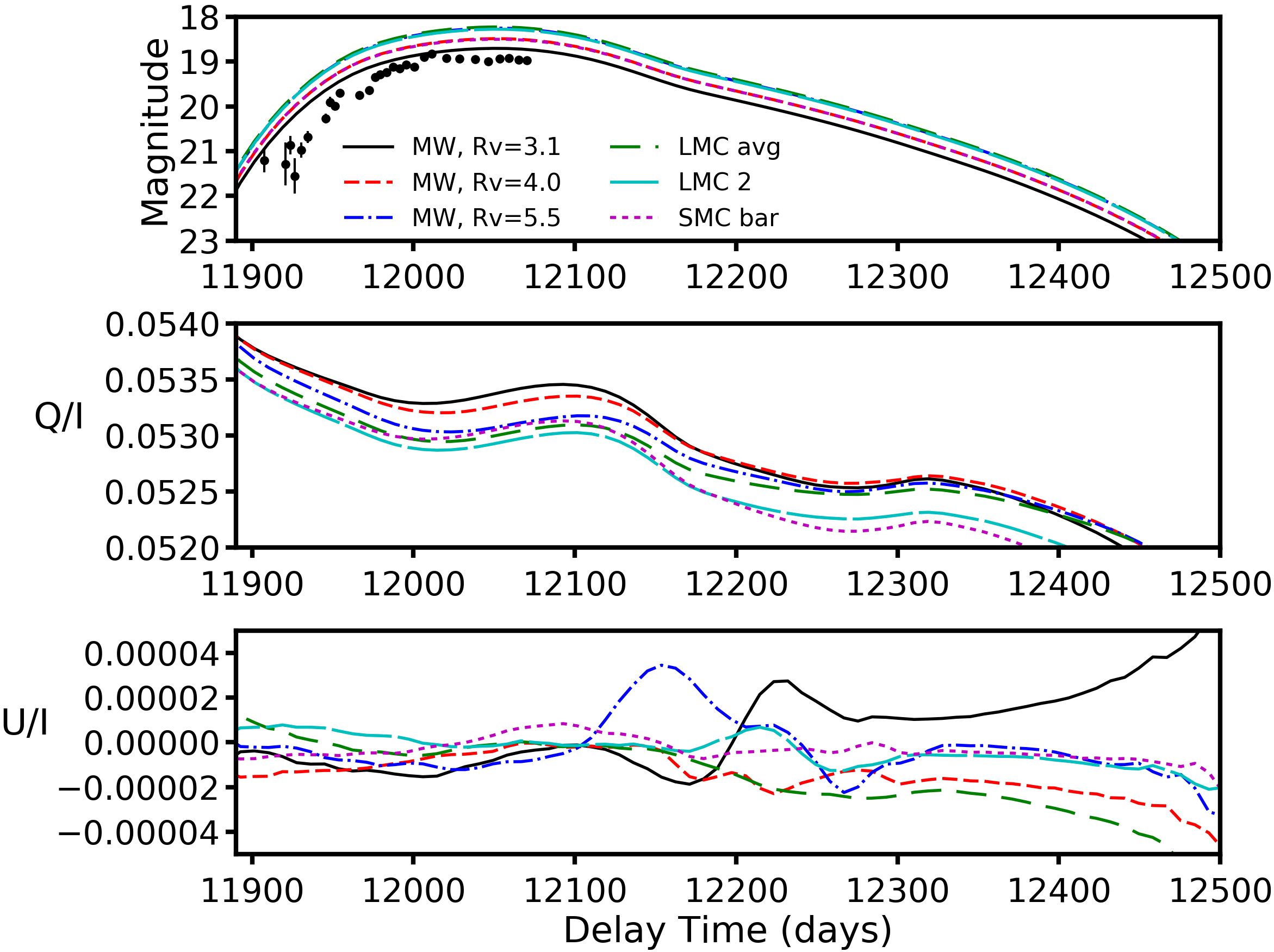}
\caption{The same as Figure \ref{fig:lc1} but with varied dust models. The dust cloud has diameter optical thickness at 0.8 ${\mu}m$ wavelength 1.0 and diameter 1.8 ly. \label{fig:lc3}}
\end{figure}

\begin{figure}[ht!]
\plotone{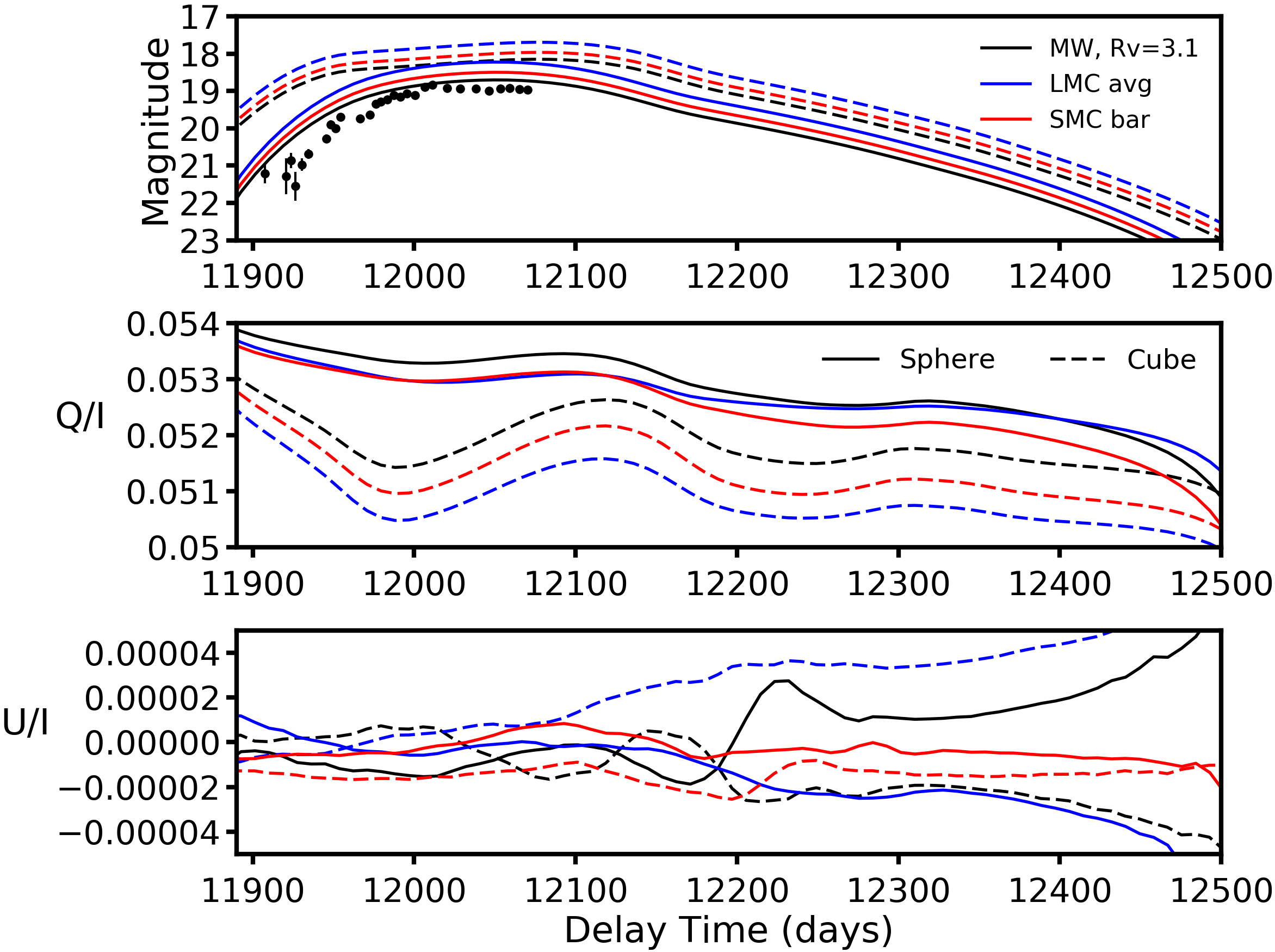}
\caption{The same as Figure \ref{fig:lc1}, but with three dust models and two dust cloud shapes. The spherical dust cloud, of 1.8 ly diameter, has optical thickness 1.0 along a diameter at 0.8 ${\mu}m$ wavelength. The cube-shaped dust cloud, with side length 1.8 ly, has optical thickness 1.0 along a side at 0.8 ${\mu}m$ wavelength. Solid lines denote spherical dust clouds and dashed lines denote cube-shaped dust clouds.\label{fig:lc4}}
\end{figure}

We use the MCRTM to simulate light echo light curves of SN~1987A at the AT2019xis position under a variety of dust cloud properties including optical thickness, dust grain model, dust cloud size and dust cloud shape. The dust cloud is assumed to be homogeneous so the extinction coefficient in Equation \eqref{eq:taus} is a constant. The simulated light curves are shown in Figures \ref{fig:lc1}-\ref{fig:lc4}.

In Figure \ref{fig:lc1}, only optical thickness is a variable. The dust cloud is assumed to be a sphere with diameter 1.8 ly and is composed of “LMC avg” dust properties. The simulated light curves can roughly capture the observed light curve shape when the diameter optical thickness is smaller than 4. The light curve maxima of the optically thin cases ($\tau<4$) arrive later than the optically thick cases. Because the dust grains are absorptive, when the dust cloud is optically thick, only areas close to the cloud surface have light come to the observer, so after the light curve maximum, the light echo luminosity decreases fast. In an optically thin cloud, the light scattered deeply into the cloud can come out of the surface and be observed, so after the light curve maximum, the light echo luminosity decreases slowly. At the same delay time, the light echo luminosity increases with increased optical thickness up to 2, and then decreases with further increased optical thickness. When the dust cloud is optically thin, if optical thickness increases moderately, more light is scattered to the observer. However, if optical thickness increases to a value where absorption is significant, light scattered to the observer would decrease. Thus, the light echo luminosity does not increase monotonically with increased optical thickness.

The linear polarization degree $p_l$ and angle $\chi$ are defined as
\begin{equation} \label{eq:pl}
p_l=\frac{\sqrt{Q^2+U^2}}{I},
\end{equation}
\begin{equation} \label{eq:chi}
\chi=\frac{1}{2}\arctan{\left(\frac{U}{Q}\right)},
\end{equation}

\noindent
where $\chi$ is the angle between the linear polarization plane and a reference plane. As shown in Figures \ref{fig:lc1}-\ref{fig:lc4}, the $Q$ component of the light echo’s Stokes vector is always positive and the $U$ component is close to zero, which indicates that the preferential linear polarization direction is parallel to the polarization reference plane, the $xz$-plane in Figure \ref{fig:obsg}. If the dust cloud and SN~1987A are treated as points, the scattering plane is the $yz$-plane. Thus, the preferential linear polarization direction is perpendicular to the scattering plane. The linear polarization degree decreases with increased optical thickness.

Figure \ref{fig:lc2} shows the effect of dust cloud diameter on the light echo. When the optical thickness is held constant, a larger dust cloud has a stronger light echo. Because a larger dust cloud receives more light from the source, more light is scattered to the observer if the optical thickness is unchanged. The linear polarization is not sensitive to the dust cloud size. As shown in Figure \ref{fig:lc2}, the variation of linear polarization is smaller than 0.1$\%$.

Figure \ref{fig:lc3} shows that the dust cloud light echo is sensitive to dust optical properties. “MW, $R_\text{V}$=5.5”, “LMC avg” and “LMC 2” have similar light echo light curves. These three dust models have similar SSAs and asymmetry factors in the I band. “MW, $R_\text{V}$=4.0” and “SMC bar” also have very similar SSAs and asymmetry factors, and their light curves are almost identical. The simulated linear polarization is not very sensitive to dust optical properties, since we have assumed that the six models have the same polarization property in Section \ref{subsec:dust}.

In Figure \ref{fig:lc4}, we compare light curves of two dust cloud shapes, namely a sphere and a cube. The shapes of light curves by a spherical and a cube-shaped dust cloud are significantly different before the light curve maxima. The light curve before the maximum of a cube-shaped dust cloud is like a line segment whereas of a spherical dust cloud has obvious curvature. The linear polarization of the two dust cloud shapes is also different, though not substantially.

\begin{figure}[ht!]
\plotone{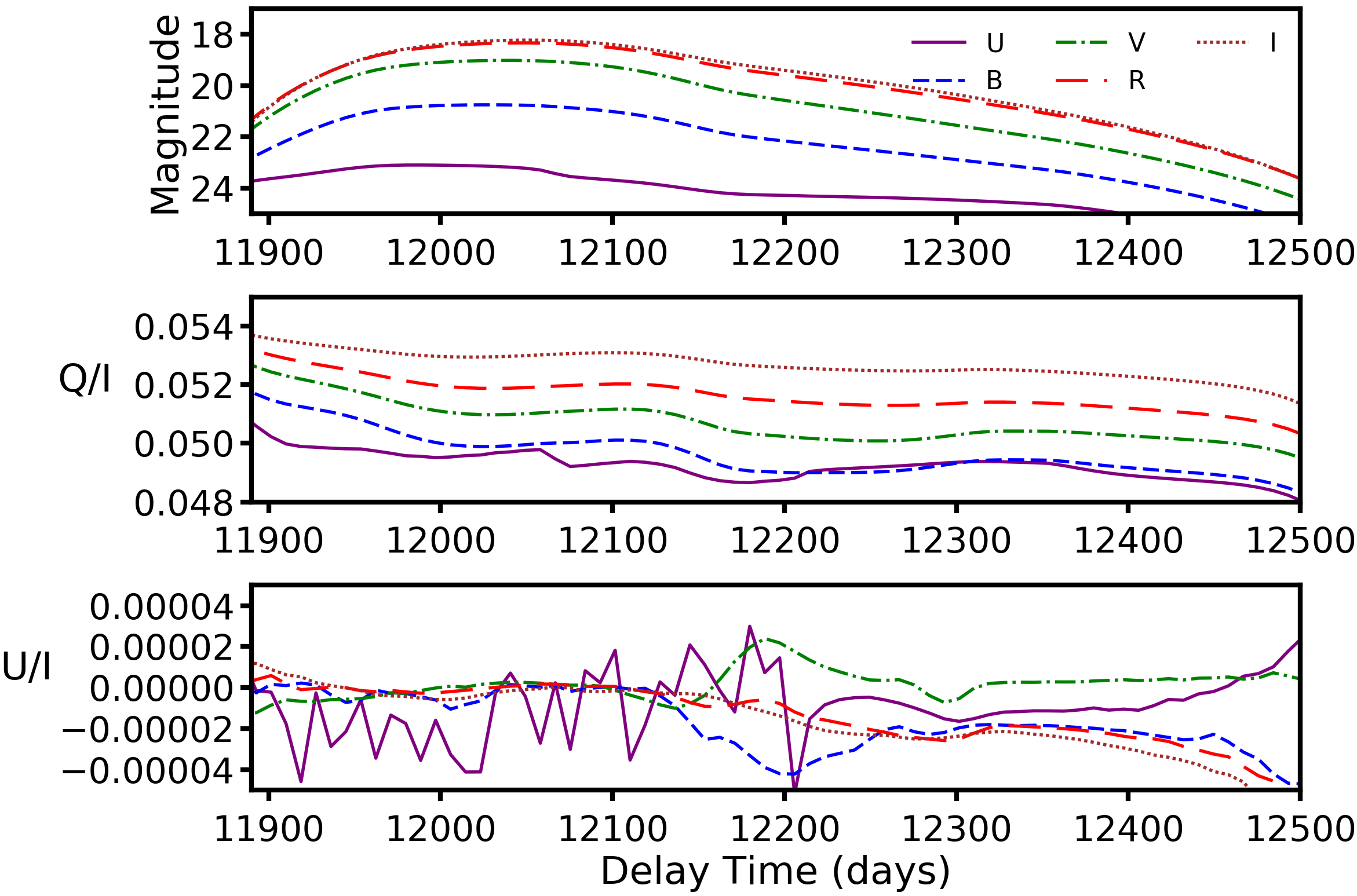}
\caption{The same as Figure \ref{fig:lc1}, but for different spectral bands (the same as in Figure \ref{fig:SN87A}). The spherical dust cloud with diameter 1.8 ly has optical thickness 1.0 along a diameter at 0.8 ${\mu}m$ wavelength (I band), increasing to about 3.2 at 0.365 ${\mu}m$ (U band). The “LMC avg” dust model is assumed.\label{fig:lc5}}
\end{figure}

We also simulate light echo light curves of the dust cloud in the U, B, V and R bands, where the Johnson-Cousins filters \citep{Landolt2009} are used. The optical thickness is 1.0 along a diameter at 0.8 ${\mu}m$ wavelength and the U, B, V and R band optical thicknesses are larger according to Figure \ref{fig:dustp} and Equation \eqref{eq:taulambda}. The simulated U, B, V, R and I band light curves are shown in Figure \ref{fig:lc5}. Besides the optical thickness, the input SN~1987A light curves and dust optical properties also vary greatly in these five bands, so the light echo light curves have a strong band dependence.

\subsection{Simulated Light Echo Images} \label{subsec:leimg}

\begin{figure*}[ht!]
\gridline{\fig{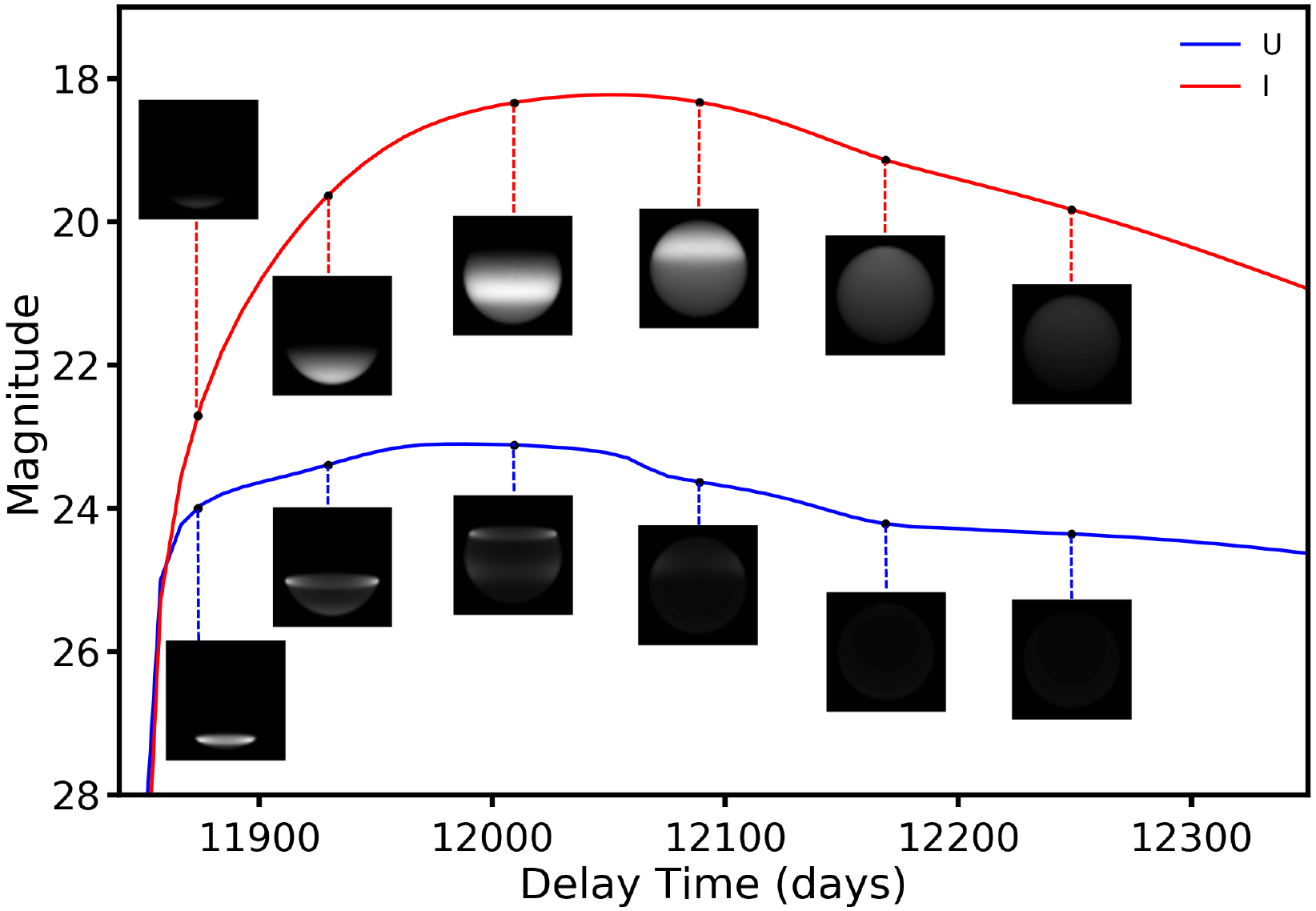}{0.6\textwidth}{(a)}}
\gridline{\fig{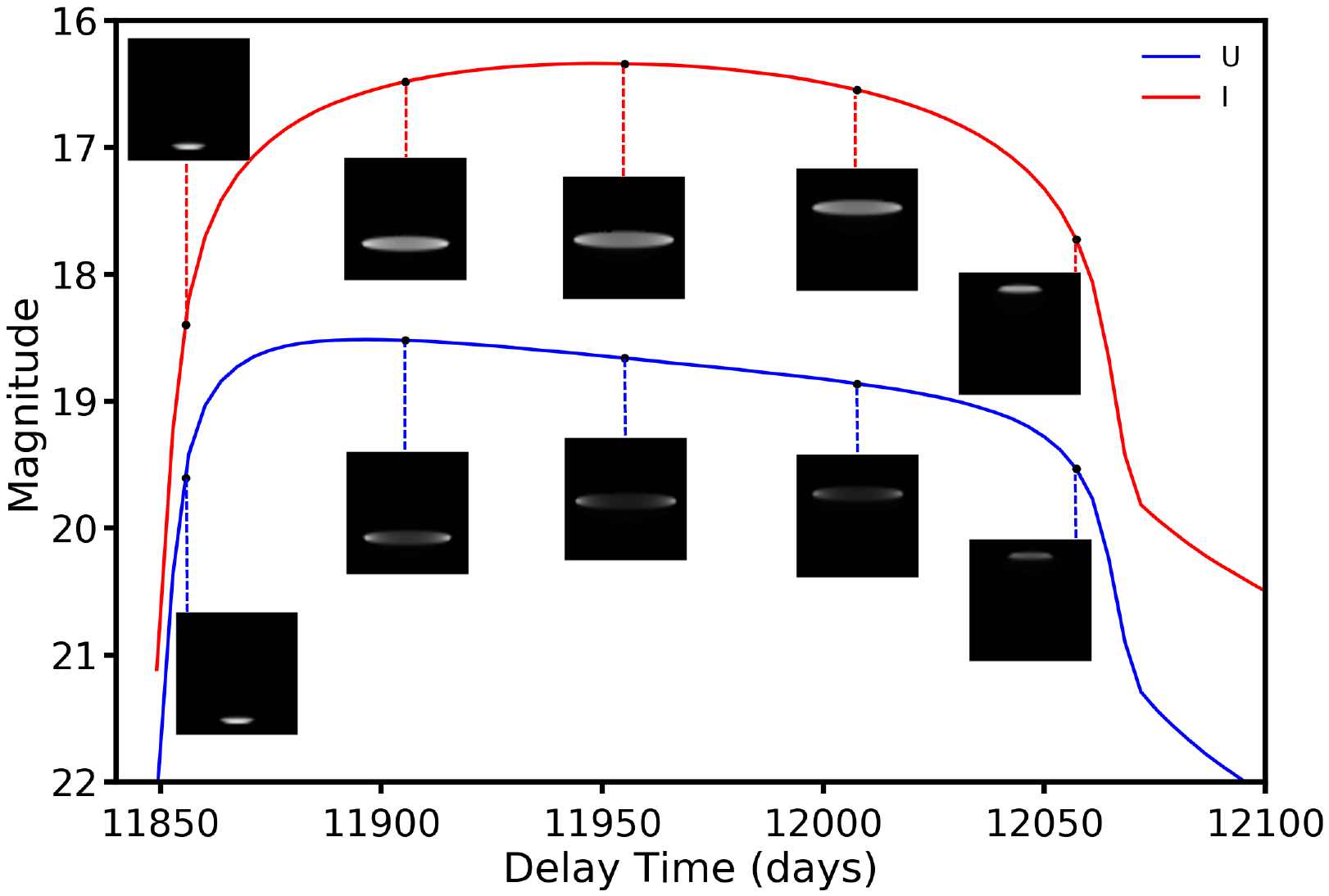}{0.6\textwidth}{(b)}}
\caption{Simulated SN~1987A I-band and U-band light echo light curves and images at several delay times. The dust cloud is assumed to be a sphere having diameter 1.8 ly. The diameter optical thickness at 0.8 ${\mu}m$ wavelength is 1.0. The “LMC avg” dust model is used in the simulation. For either band, the same colormap is used to plot all grayscale images in this band, but the colormaps used for I and U bands are different. \textbf{(a)}: SN~1987A light curve is used in the simulation; \textbf{(b)}: A Dirac Delta function multiplied by the maximum flux of SN~1987A photometry is the supernova’s light curve in the simulation.\label{fig:img1}}
\end{figure*}

\begin{figure*}[ht!]
\gridline{\fig{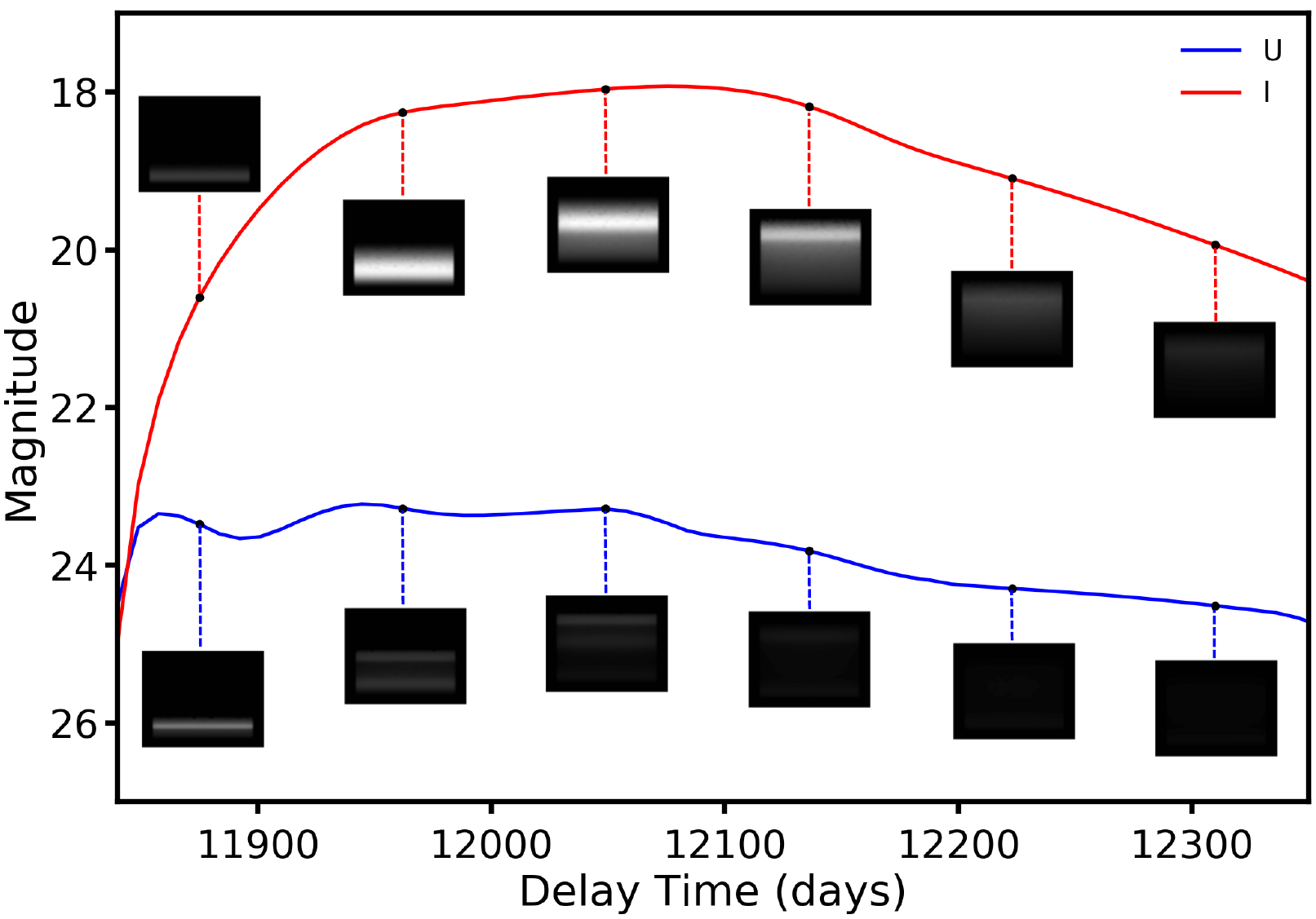}{0.6\textwidth}{(a)}}
\gridline{\fig{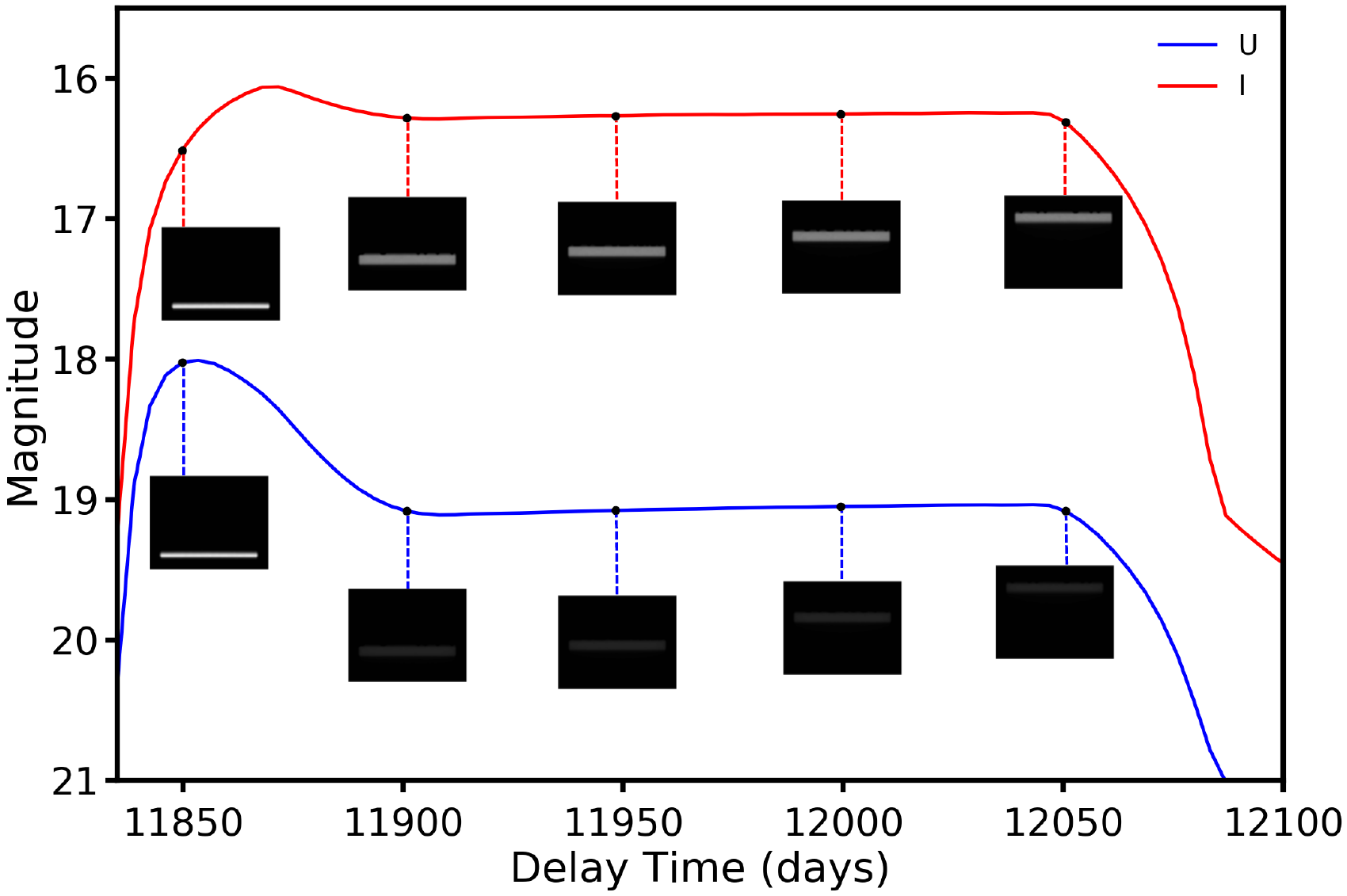}{0.6\textwidth}{(b)}}
\caption{The same as Figure 13 but for a cube shape. The side length is 1.8 ly. The side optical thickness at 0.8 ${\mu}m$ wavelength is 1.0.\label{fig:img2}}
\end{figure*}

In this subsection, we show the light echo images simulated by the MCRTM. For illustrative purposes, we also show the impulse response of the corresponding scattering event, where the supernova’s light curve is modeled by a Dirac Delta function. Figures \ref{fig:img1}-\ref{fig:img2} show the simulated U-band and I-band light echo images at different delay times. The optical thickness in the I band is small (1.0 at 0.8 ${\mu}m$ wavelength). The corresponding optical thickness at 0.365 ${\mu}m$ is about 3.2 so the U band has a larger optical thickness.

As shown in Figures \ref{fig:img1}-\ref{fig:img2}, after the I-band light curve maximum, the whole dust cloud is illuminated for a while. A bright horizontal strip is caused by the SN~1987A I-band light curve maximum as shown in Figure \ref{fig:SN87A}. In the corresponding impulse response, the bright horizontal strip is narrow and there is almost no light outside the strip. According to the analysis using Figure \ref{fig:view2D}, the single scattering contribution is very significant. The bright strip first appears at the bottom of the dust cloud, then moves upward with time and finally disappears at the top. The movement of the bright strip is consistent with the upward moving light echo paraboloid. Because of the position of the dust cloud and SN~1987A, for a spherical dust cloud, the light observed at the top of the cloud experiences a longer distance within the dust cloud than at the bottom and thus is dimmer. This phenomenon is very obvious in the I-band images in Figure \ref{fig:img1}. The light curve maximum appears at the time when the strip just passes the center of the dust cloud. In contrast, for a cube-shaped dust cloud, the singly scattered light echo has the same travel distance in the dust cloud. Thus, the bright strip of the light echo has almost the same brightness when moving from the bottom to the top of the dust cloud.

Because the U-band and I-band light curve maxima of SN~1987A are almost synchronous, a bright strip appears in the same position in the U-band image as in the corresponding I-band image. In addition to the bright strip coinciding with that in the I-band image, a much brighter and narrower strip appears ahead of the dimmer strip, which is very similar to the bright strip of the corresponding impulse response. This brighter strip is caused by the shock breakout following shortly after the SN~1987A explosion. The U-band light curve of SN~1987A in Figure \ref{fig:SN87A} shows the fading of the UV-bright shock breakout, which is strong and lasts a very short time so the light echo caused by the UV shock breakout is close to an impulse response. In the U band, the optical thickness is large. For a spherical dust cloud, most of the observed light is concentrated close to the dust cloud surface since that region experiences less extinction. In the U-band light echo image in Figure \ref{fig:img1}, the bright strip is dim in the center and bright at its two ends. This inhomogeneous bright strip brightness is not shown in the U-band light echo image for a cube-shaped dust cloud, because the light has the same extinction across its bright strip.

The U-band light curve is flatter than the I-band light curve around the maximum as shown in Figures \ref{fig:img1}-\ref{fig:img2} and Figure \ref{fig:lc5}. The impulse response light curves and images in the lower plots of Figures \ref{fig:img1}-\ref{fig:img2} can explain the U-band flat light curves. In the U band, the light echo of the strong UV-bright shock dominates the received signal so its light curve shape resembles to the corresponding impulse response, which is flat when the bright strip moves from one side of the dust cloud to another side. For an impulse response, the magnitude of the light echo is proportional to the dust volume that the light echo paraboloid intersects.

In the lower plots of Figures \ref{fig:img1}-\ref{fig:img2}, the area and brightness of the bright strip determine the light echo magnitude. In other words, the impulse response light curve depicts the shape of the dust cloud. For a spherical dust cloud, the I-band impulse response light curve is almost symmetric with respect to the maximum since the I-band extinction is weak. In contrast, the U-band impulse response light curve looks like a symmetric light curve multiplied by an attenuation term that increases with the delay time. Based on the left plot of Figure \ref{fig:view2D}, when the bright strip moves from the bottom to the top of the spherical dust cloud, the average pathlength of photons in the dust cloud is longer so the dust extinction is stronger in both U and I bands. The U-band optical thickness is larger than in the I band, so the decreasing trend in the U-band light curve is more obvious. For a cube-shaped dust cloud, the I-band and U-band impulse response light curves have similar shapes. The light echo magnitude gradually reaches its maximum in the stage where the width of the bright strip increases. Then, the light curve decreases to a point and becomes flat with a slight increasing trend. In this stage, the width of the bright strip is constant and the pathlength of singly scattered photons in the dust cloud at various delay times is the same. The slight increasing trend is attributed to multiple scattering. Later on, the impulse response light curve starts to decrease, which coincides with the decreasing bright strip width. In short, the light curve shape of the impulse response is a combined result of dust cloud shape and extinction.

\begin{figure}[ht!]
\plotone{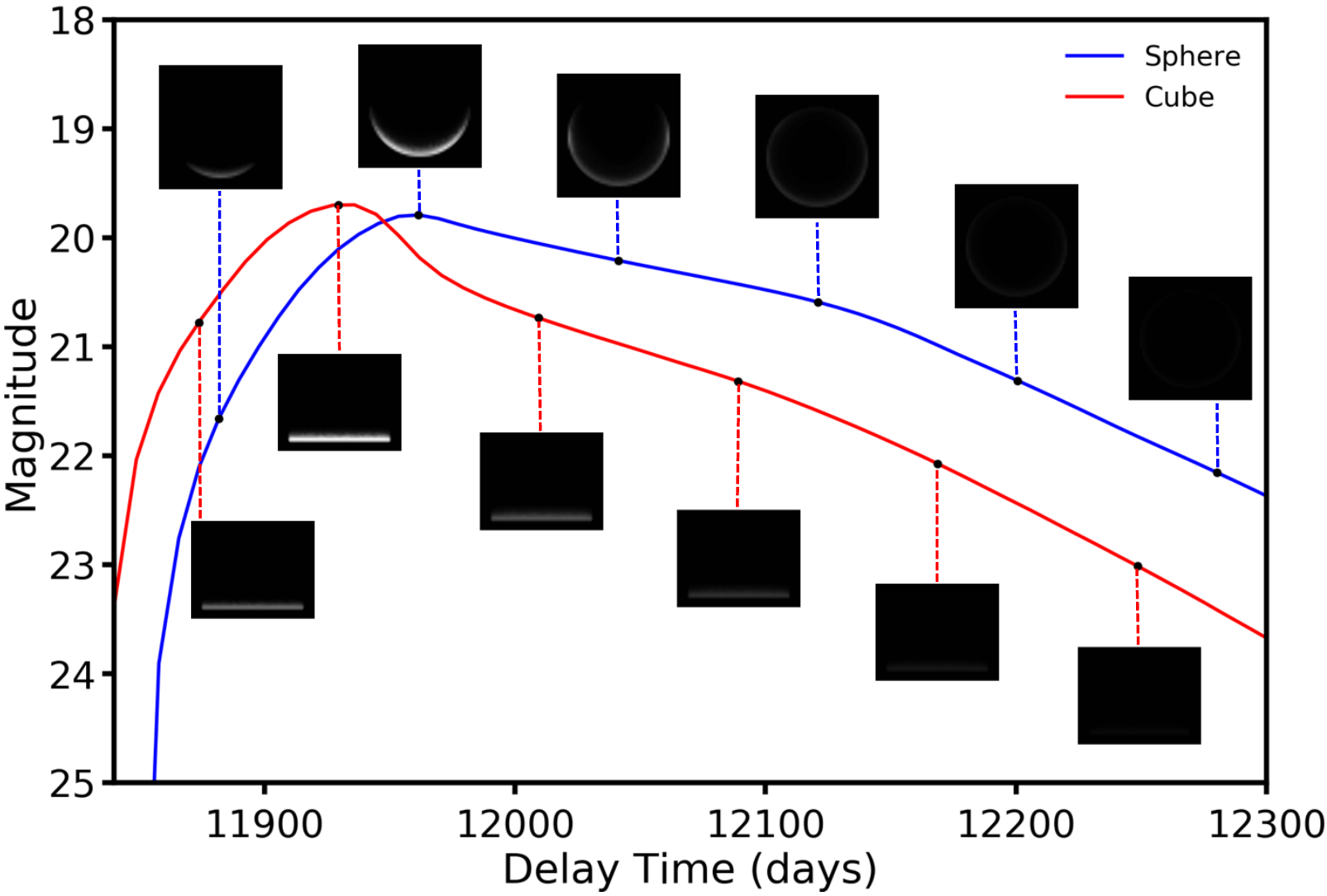}
\caption{Simulated SN~1987A I-band light echo light curves and images at six delay times for a spherical and a cube-shaped dust cloud shape. The spherical dust cloud has diameter 1.8 ly and diameter optical thickness 10.0 at 0.8 ${\mu}m$ wavelength. The cube-shaped dust cloud has side 1.8 ly and side optical thickness 10.0 at 0.8 ${\mu}m$ wavelength. The “LMC avg” dust model is used in the simulation. For each dust cloud shape, the same colormap is used to plot all grayscale images. The colormaps used for spherical and cube-shaped dust clouds are different.\label{fig:img3}}
\end{figure}

Figure \ref{fig:img3} shows the simulated I-band light echo images when the dust cloud is optically thick. For a spherical dust cloud, the light echo looks like an incomplete ring since only the light scattered by the sphere’s surface is observable. At the light curve maximum, the incomplete ring is less than 180$^{\circ}$ when the SN~1987A I-band light curve maximum has not passed the center of the sphere yet. This explains why the light curve maximum of an optically thick cloud appears earlier than an optically thin cloud. For a cube-shaped dust cloud, only the light scattered by the bottom surface is observable. Because a cube is a faceted shape and four faces are parallel to the incident light, the left and right surfaces of the cube-shaped dust cloud in Figure \ref{fig:img3} cannot scatter light toward the observer. The regions adjacent to the left and right surfaces are optically thick along the observation direction. Therefore, the observer can only see the light echo at the cube’s bottom region.

\begin{figure}[ht!]
\plotone{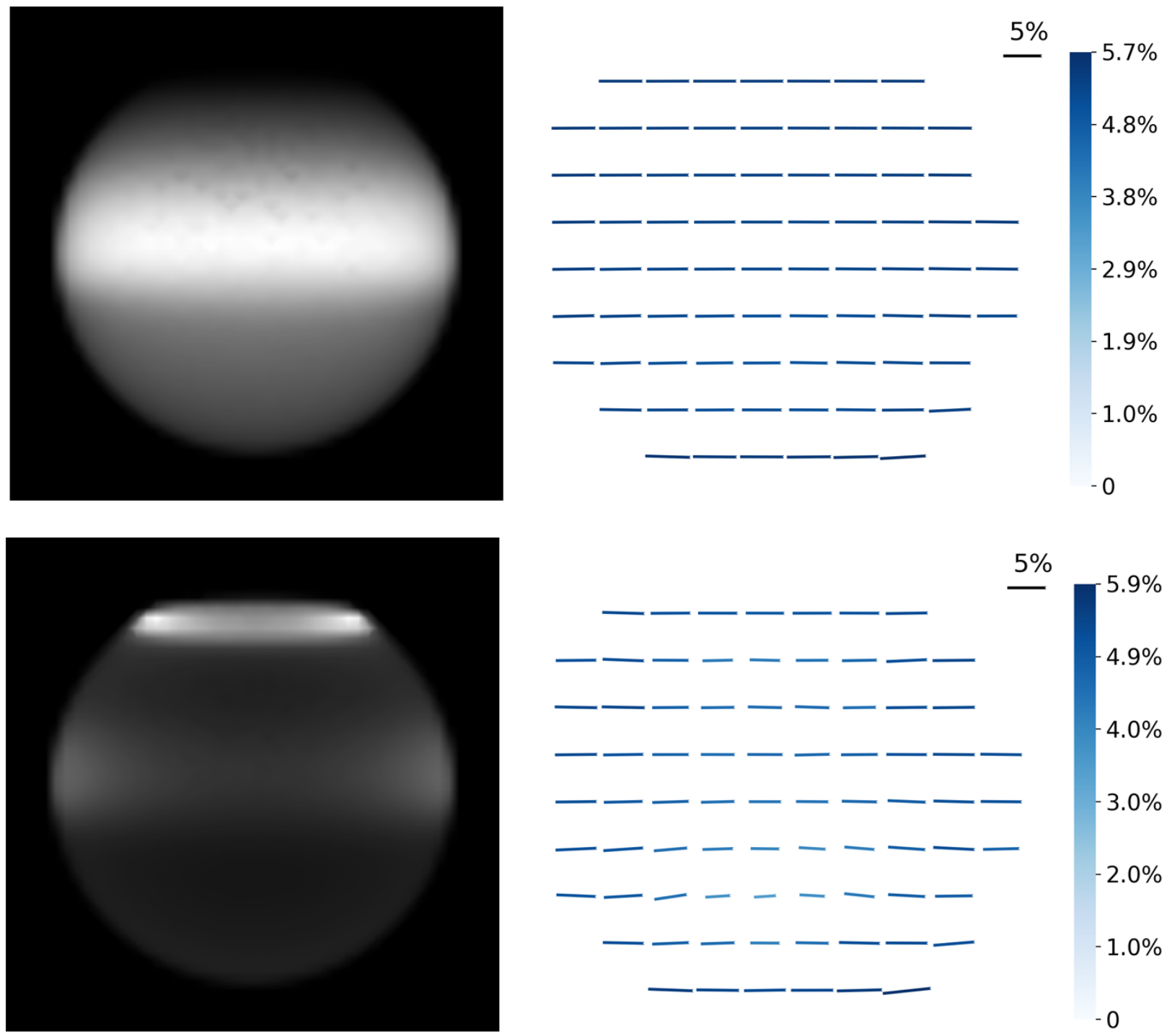}
\caption{Simulated SN~1987A I-band and U-band light echo images and linear polarization patterns. The dust cloud is the same as in Figure \ref{fig:img1}. \textbf{Top row}: I band; \textbf{Bottom row}: U band. \textbf{Left column}: simulated light echo images; \textbf{Right column}: simulated linear polarization patterns. The length and color of the line segments in linear polarization patterns indicate the linear polarization degree defined in Equation \eqref{eq:pl}. The ‘5\%’ marker denotes linear polarization degree 5\%. The pointing direction of a line segment forms an angle with the right horizontal direction, which is the polarization angle defined in Equation \eqref{eq:chi}. The two grayscale images use different colormaps.\label{fig:pol1}}
\end{figure}

\begin{figure}[ht!]
\plotone{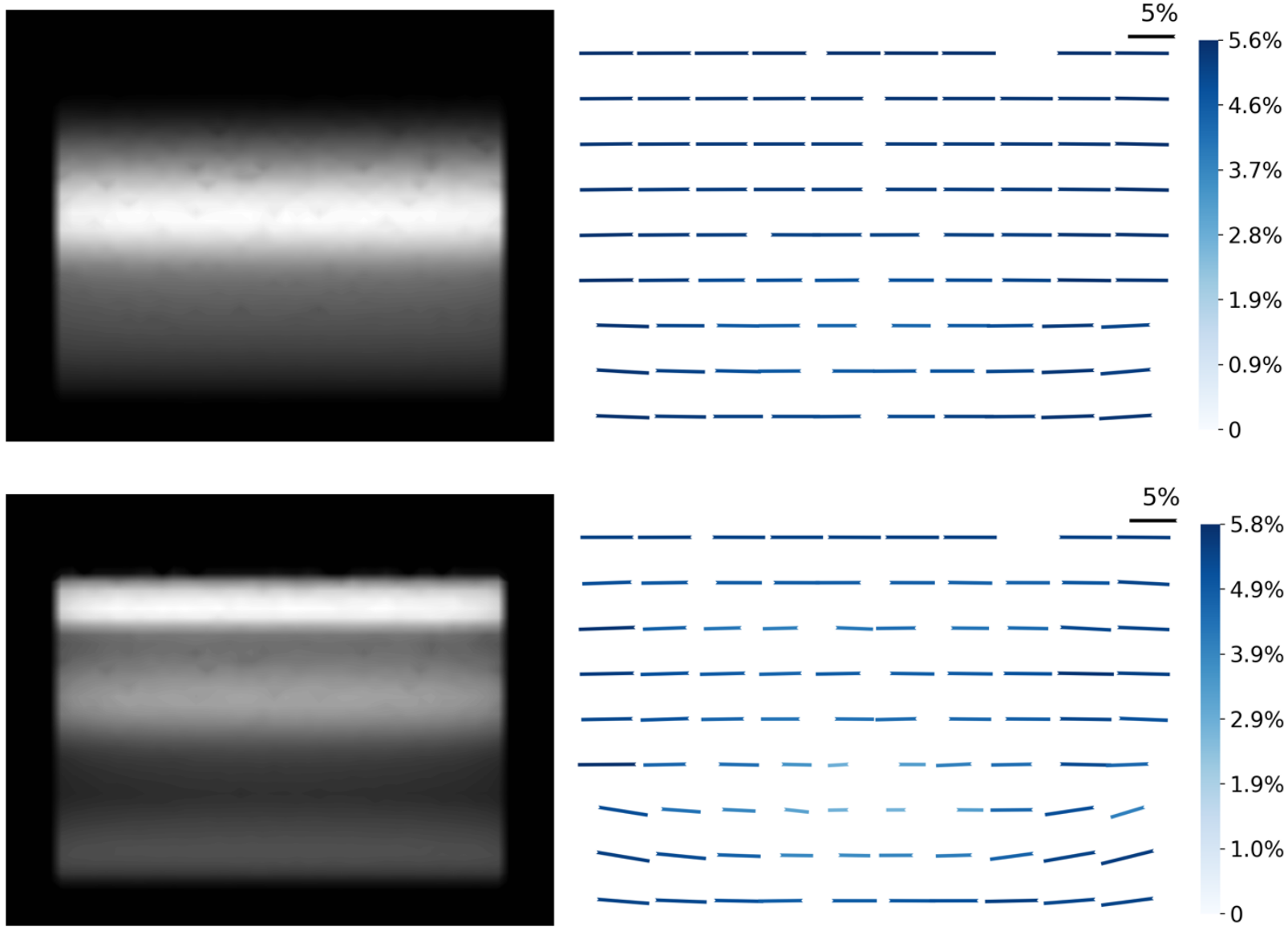}
\caption{The same as Figure \ref{fig:pol1}, but for dust cloud defined in Figure \ref{fig:img2}.\label{fig:pol2}}
\end{figure}

\begin{figure}[ht!]
\plotone{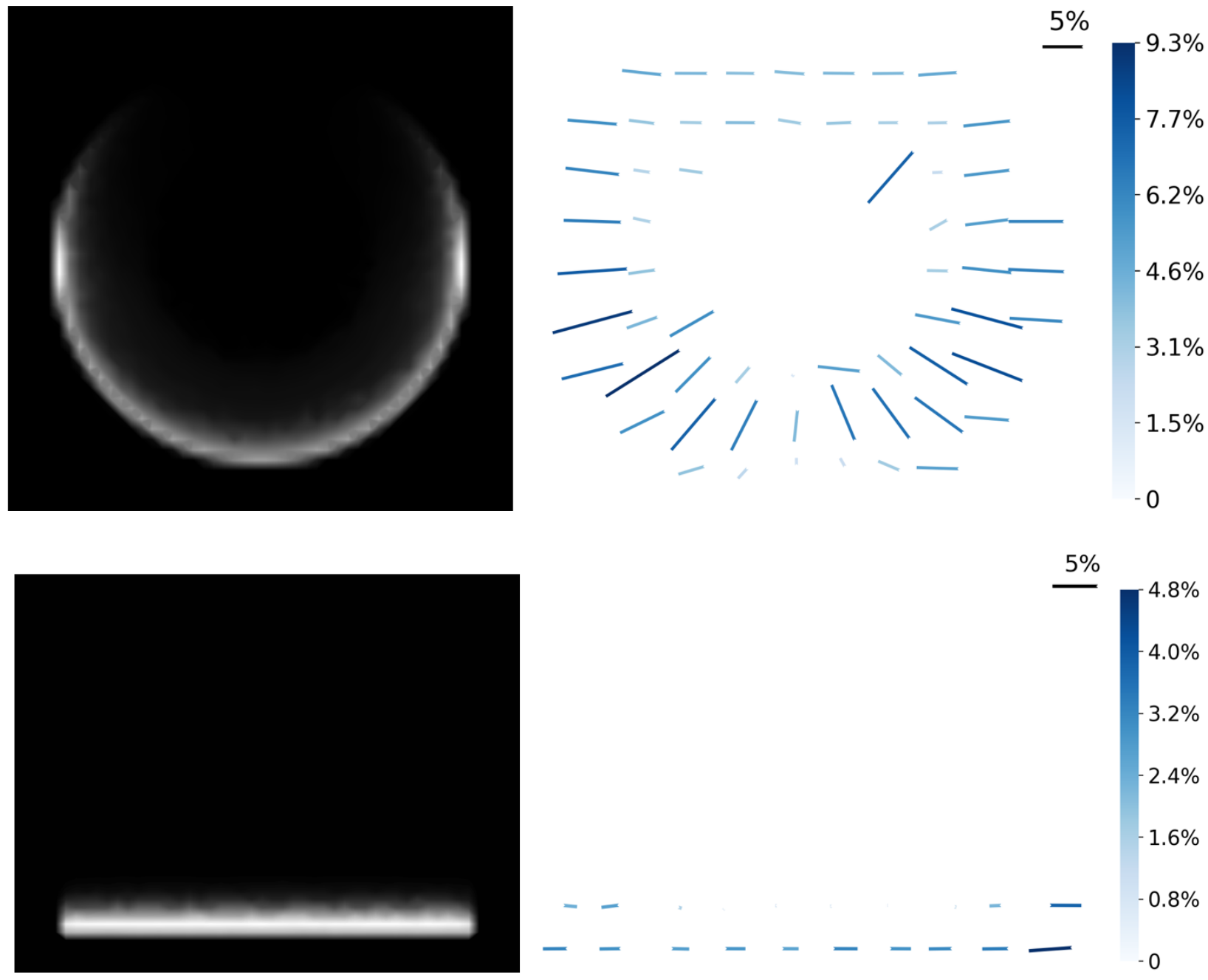}
\caption{Simulated SN~1987A I-band light echo images and linear polarization patterns. The two dust clouds are the same as in Figure \ref{fig:img3}. \textbf{Top row}: spherical dust cloud; \textbf{Bottom row}: cube-shaped dust cloud. \textbf{Left column}: simulated light echo images; \textbf{Right column}: simulated linear polarization patterns. The length and color of the line segments in linear polarization patterns indicate the linear polarization degree defined in Equation \eqref{eq:pl}. The ‘5\%’ maker denotes a linear polarization degree 5\%. The pointing direction of a line segment forms an angle with the right horizontal direction, which is the polarization angle defined in Equation \eqref{eq:chi}. The two grayscale images use different colormaps.\label{fig:pol3}}
\end{figure}

Figures \ref{fig:pol1}-\ref{fig:pol3} give zoomed views and linear polarization patterns at a single delay time of the light echo images in Figures \ref{fig:img1}-\ref{fig:img3} respectively. In Figures \ref{fig:pol1}-\ref{fig:pol2}, the light echo illuminates the whole dust cloud. The linear polarization pattern does not vary substantially in an I-band light echo. The linear polarization degree is around 5\% and the preferential polarization direction is generally perpendicular to the $yz$-plane. In a U-band light echo, the linear polarization degree around the center is smaller than that adjacent to the surface. In some areas, the preferential polarization direction is not perpendicular to the $yz$-plane, which indicates the U Stokes parameter values in these areas are nonzero.

Because the dust clouds in our simulations are homogeneous and symmetrical about the incident direction, the simulated light echoes are also symmetrical about the incident direction if the Monte Carlo noise is excluded. The U Stokes parameter of a point at one side of the axis of symmetry should have the same magnitude but opposite sign as its symmetric point in another side of the axis of symmetry. Therefore, as shown in Figures \ref{fig:lc1}-\ref{fig:lc5}, the averaged U Stokes parameters are nearly zero, although they may not be zero everywhere on the light echo.

For the optically thick case in Figure \ref{fig:pol3}, most unilluminated regions have zero polarization. In illuminated regions of the spherical dust cloud, the preferential polarization direction is radial, which is perpendicular to the local scattering plane that is parallel to the sphere surface. Because the dust cloud is optically thick, observable light must go through a very short path in the dust. Only light scattered by the dust cloud surface satisfies the short-path condition. Because almost all the observed light goes through a very short path in the dust, singly scattered light dominates the observation. Therefore, the linear polarization degrees in some regions are up to 10\% and the preferential polarization directions are along radial directions. The top of the spherical dust cloud is dim but has horizontally distributed linear polarization, which is weaker than the radially distributed counterpart below. This horizontally distributed linear polarization is dominated by multiple scattering, which tends to reduce polarization.

\subsection{Estimation of Dust Cloud Size and Optical Thickness} \label{subsec:est}

\begin{figure}[ht!]
\plotone{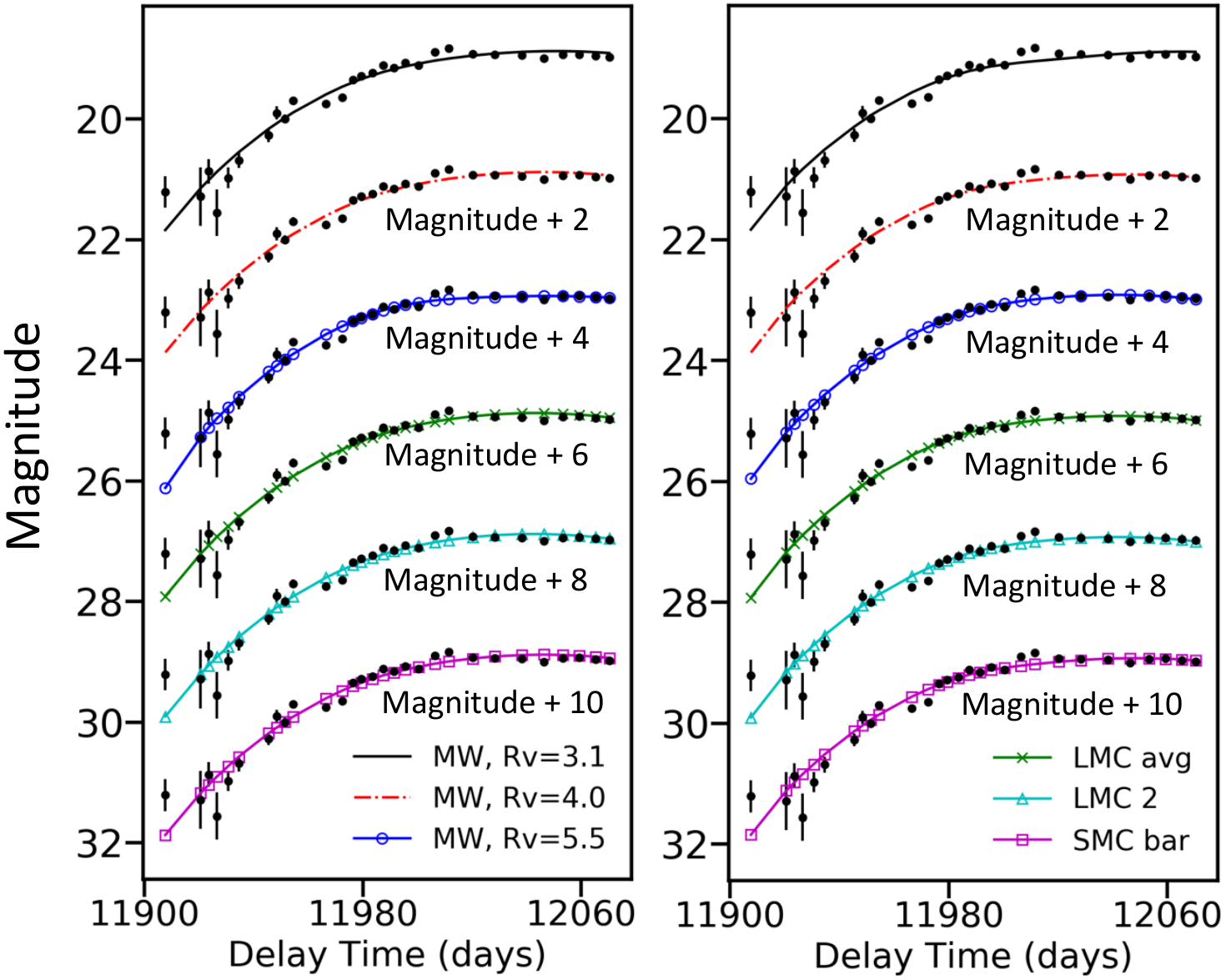}
\caption{Best-fit simulated I-band light echo light curves to the OGLE-IV OTDS AT2019xis observation using six dust models. The black dots are AT2019xis I-band observations. \textbf{Left}: The dust cloud is assumed to be a sphere; \textbf{Right}: The dust cloud is assumed to be a cube. The magnitudes of “MW, $R_\text{V}$=4.0”, “MW, $R_\text{V}$=5.5”, “LMC avg”, “LMC 2” and “SMC bar” are shifted for clarity.\label{fig:est}}
\end{figure}

We use the MCRTM simulations to fit the OGLE-IV OTDS I-band light curve and estimate dust cloud size and optical thickness. The best-fit light curves are shown in Figure \ref{fig:est}. Table \ref{tab:est} lists the estimated dust cloud size and optical thickness.

All the best-fit light curves fit the overall shape of the observed light curve well but none of them captures the local minima and maxima before and after delay time 11980 days. Our simulations assume that the dust cloud has simple and regular shapes but the realistic dust cloud is irregular and inhomogeneous, as shown in Figure \ref{fig:hstimg}. The complex shape and inhomogeneity result in the observed light curve’s local minima and maxima.

After delay time 11980 days, the light curves gradually become flat. After delay time 12060 days, the observed light curve tends to decrease. Because we do not have more observations, we are not sure if this decrease is a local minimum or a decreasing trend of the light curve after a global maximum. As shown in Table \ref{tab:est}, the estimated dust cloud size and optical thickness by “MW, $R_\text{V}$=3.1”, “MW, $R_\text{V}$=4.0” and “SMC bar” dust models are larger than the two LMC dust models. The “MW, $R_\text{V}$=5.5” model has larger optical thickness but smaller size than the LMC dust models. We find in subsection \ref{subsec:lcs} that, at the same delay time, the light echo magnitude is larger for a larger dust cloud if optical thickness is unchanged. For any dust cloud size and the same delay time, there is a maximum achievable light echo magnitude. The maximum achievable light echo magnitude is determined by the dust cloud absorption capability and size. A more absorptive dust cloud scatters less light and therefore has a smaller maximum achievable light echo magnitude. The “MW, $R_\text{V}$=3.1”, “MW, $R_\text{V}$=4.0” and “SMC bar” models have smaller I-band SSAs and thus are more absorptive than other three models. Thus, to generate a light curve fitting the observed light curve, the dust cloud size of “MW, $R_\text{V}$=3.1”, “MW, $R_\text{V}$=4.0” and “SMC bar” models must be larger than those of other models.

The mass of the dust cloud can be computed using the estimated cloud size and optical thickness,
\begin{equation} \label{eq:masss}
M=\frac{{\pi}D^2{\tau}_D}{6C_\text{ext}},
\end{equation}

\noindent
for a spherical dust cloud, and
\begin{equation} \label{eq:massc}
M=\frac{a^2{\tau}_a}{C_\text{ext}},
\end{equation}

\noindent
for a cube-shaped dust cloud. $D$ is diameter and ${\tau}_D$ is the optical thickness along a sphere diameter. $a$ is side length and ${\tau}_a$ is the optical thickness along a cube side. The estimated mass data are listed in Table \ref{tab:est}. Except for “MW, $R_\text{V}$=5.5” model, the estimated mass from assuming a spherical shape is slightly smaller than that by assuming a cube shape if the same dust model is used, which indicates that dust cloud shape assumption does not significantly affects the mass estimation result.

If we assume a gas-to-dust ratio of 300, which is within the reported gas-to-dust ratio range in the LMC (e.g., \citealt{Roman-Duval2014}), and use the minimum and maximum estimated dust masses with the two LMC dust models, the total mass of the cloud is about 7.8-9.3 \(M_\odot\). If we consider either homogeneous spherical or cube-shaped cloud, the cloud mass is estimated to be 7.8-8.4 \(M_\odot\) or 8.1-9.3 \(M_\odot\) respectively. If we assume the dust has MW dust optical properties, the estimated mass is 7.2-29.0 \(M_\odot\) and 8.4-16.3 \(M_\odot\) for spherical and cube-shaped clouds respectively. The dust optical property uncertainty is comparable to the cloud shape uncertainty. Furthermore, the estimated mass ranges have similar smallest values but very different largest values. This implies that according to radiative transfer calculation, the scattered flux at I band is not sensitive to optical thickness variation when optical thickness is large.

\begin{deluxetable*}{lcccc}
\tablenum{1}
\tablecaption{Best-fit dust cloud size, optical thickness ($\tau$) and estimated mass\label{tab:est}}
\tablewidth{0pt}
\tablehead{
\colhead{Dust Model} & \colhead{Shape} & \colhead{Size$^*$ (ly)} &
\colhead{${\tau}^{**}$} & \colhead{Mass (\(M_\odot\))}}
\startdata
MW, $R_\text{V}$=3.1 & Sphere & 1.59 & 1.07	& 0.042 \\
MW, $R_\text{V}$=4.0 & Sphere & 1.54 & 0.74	& 0.024 \\
MW, $R_\text{V}$=5.5 & Sphere & 1.43 & 3.78	& 0.096 \\
LMC avg.             & Sphere & 1.51 & 0.50	& 0.026 \\
LMC 2.               & Sphere & 1.50 & 0.54	& 0.028 \\
SMC bar	             & Sphere & 1.54 & 0.69	& 0.044 \\
MW, $R_\text{V}$=3.1 & Cube   & 1.32 & 1.03	& 0.053 \\
MW, $R_\text{V}$=4.0 & Cube	  & 1.16 & 0.82	& 0.028 \\
MW, $R_\text{V}$=5.5 & Cube	  & 1.11 & 0.59	& 0.017 \\
LMC avg	             & Cube	  & 1.12 & 0.51	& 0.027 \\
LMC 2	             & Cube	  & 1.11 & 0.56	& 0.031 \\
SMC bar              & Cube	  & 1.17 & 0.82	& 0.057 \\
\enddata
\tablecomments{$^*$For sphere, size is the diameter; for cube, size is the cube side length.
\\$^{**}$For sphere, $\tau$ is along diameter; for cube, $\tau$ is along a cube side. All $\tau$ values are at 0.8 ${\mu}m$ wavelength.}
\end{deluxetable*}

With the best-fit dust cloud optical thickness and size, we can simulate the light echo in other unobserved bands. Here we take Swift \citep{Gehrels2004} UV/Optical Telescope (UVOT) \citep{Roming2005} as an example to simulate the light echo light curves in six UVOT filters (v, b, u, uvw1, uvw2 and uvm2). The UVOT filter effective area data and zero points in Vega system \citep{Breeveld2011} are obtained from the SVO Filter Profile Service \citep{2012ivoa.rept.1015R,2020sea..confE.182R} \footnote{http://svo2.cab.inta-csic.es/theory/fps/}.
To calculate the brightness of the light echo in the (unobserved) UV, we use UV spectra from the International Ultraviolet Explorer (IUE) published by \cite{Pun1995}. These IUE spectra were combined with optical spectra from \citet{Phillips1988,Phillips1990}. The combined spectra are available from the OSC \citep{Guillochon2017} and the Weizmann Interactive Supernova Data Repository (WISeREP; \citealp{Yaron2012}) \footnote{https://wiserep.weizmann.ac.il}.
Low or zero flux gaps at the beginning, end, or between spectra were deleted and the gaps between spectra replaced with a linear interpolation.

The combined SN~1987A spectra are at 28 epochs ranging from -81 to 632 days relative to the date of B-band maximum (Modified Julian Day 46931.0). We construct the SN~1987A light curve magnitude in the UVOT v, b, u, uvw1, uvw2 and uvm2 bands using the formulas:
\begin{subequations} \label{eq:flxavg}
\begin{equation}
\bar{F}=\frac{{\int}T(\lambda)\,F(\lambda)\,d\lambda}{{\int}T(\lambda)\,d\lambda},
\end{equation}
\begin{equation}
\mathrm{mag}=-2.5\log_{10}\left(\frac{\bar{F}}{F_0}\right),
\end{equation}
\end{subequations}
where $\bar{F}$ is band-averaged flux, $F(\lambda)$ is the flux at a single wavelength, $F_0$ is the Vega zero-point flux and $\mathrm{mag}$ is the magnitude in Vega system. With the constructed flux $\bar{F}$ at a specific filter, we can calculate the light echo brightness using Equation \eqref{eq:stkavg} and \eqref{eq:stkle}. Since we have SN~1987A UV spectra, we can calculate the light echo brightness with a more strict approach. The light echo flux is first computed wavelength by wavelength and then integrated with the filter transmission to get the band-averaged flux,
\begin{equation} \label{eq:flxle}
\bar{F}_{LE}(t)={\int}{\int}F_{\delta}(\lambda,t-t')\,F_{\text{SN}}(\lambda,t')T(\lambda)\,d{\lambda}\,dt',
\end{equation}
where $\bar{F}_{LE}$ is light echo flux at a specific band, $F_{\delta}$ is the simulated impulse response flux of the light echo and $F_{\text{SN}}$ is supernova spectral flux.

\begin{figure}[ht!]
\plotone{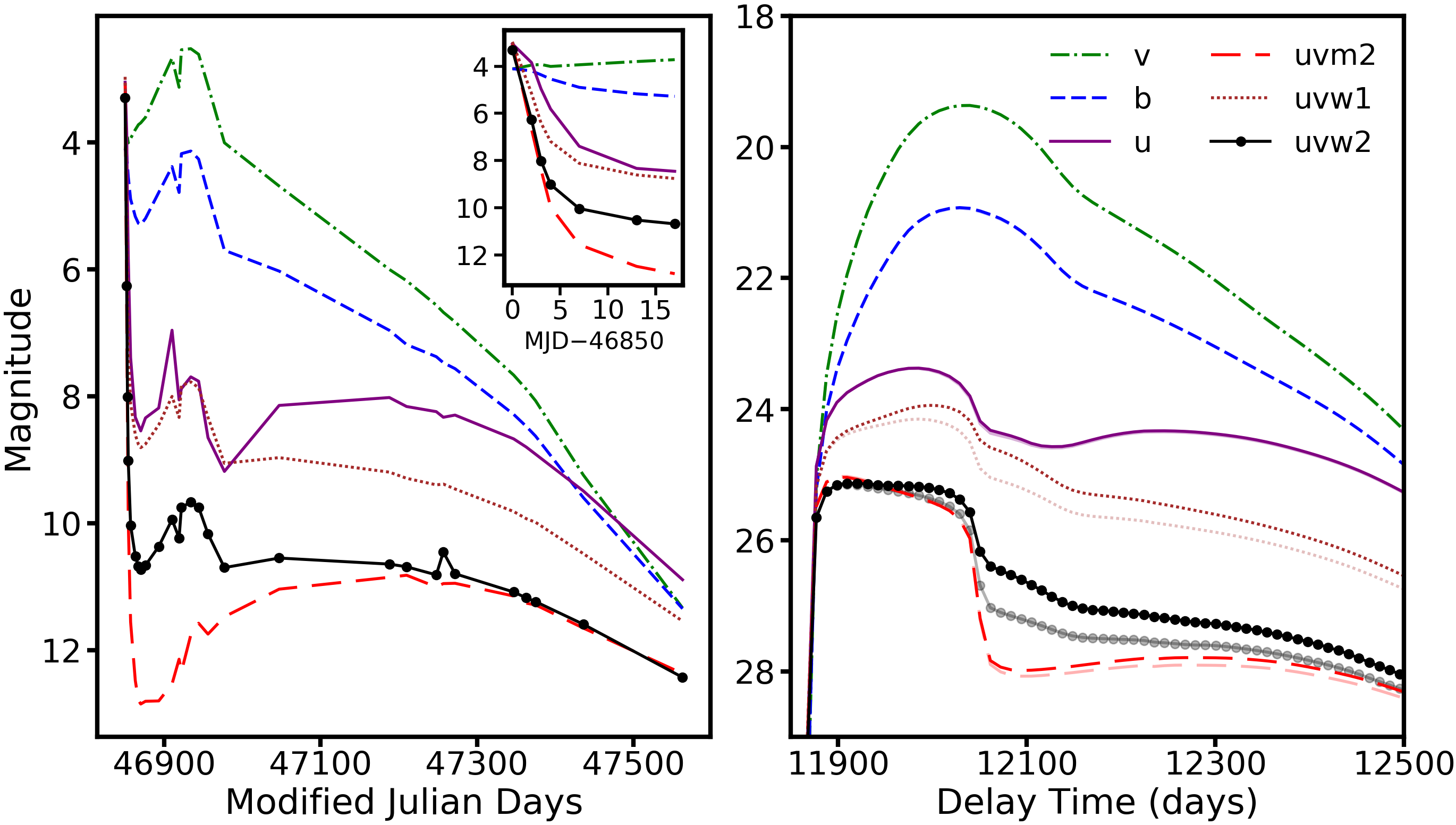}
\caption{\textbf{Left}: Constructed SN~1987A light curves in Swift UVOT v, b, u, uvw1, uvw2 and uvm2 bands using the combined SN~1987A spectra at 28 epochs; \textbf{Right}: The simulated light echo light curves in the six corresponding Swift UVOT filters. The dust cloud is assumed to be a sphere having diameter 1.51 ly. The “LMC avg” dust model is used in the simulation. The diameter optical thickness at 0.8 ${\mu}m$ wavelength is 0.5. The optical thicknesses at v, b, u, uvw1, uvw2 and uvm2 are about 0.95, 1.29, 1.65, 2.23, 3.53, and 3.45 respectively. Equation \eqref{eq:flxle} is used to compute the light echo light curves. The corresponding curves with lighter colors are light curves computed by the approximated approach using Equations \eqref{eq:stkavg} and \eqref{eq:stkle}. The inset plot is a zoomed view of SN~1987A light curve 0 to 17 days from modified Julian day (MJD) 46850. \label{fig:swiftuvot}}
\end{figure}

In the calculation, we pick the best-fit dust cloud optical thickness and size in Table \ref{tab:est} with the assumption of “LMC avg” dust model and spherical cloud shape. The results are shown in Figure \ref{fig:swiftuvot}. The simulated UVOT v and b band light curve shapes are similar to the V and B band light curves in Figure \ref{fig:lc5}. The strong and short UV-bright shock breakout makes the light echo UV light curves close to the impulse response as explained in subsection \ref{subsec:leimg}. The light echo light curves are calculated with two approaches. One is consistent with Equations \eqref{eq:stkavg} and \eqref{eq:stkle}, and another is more strict using Equation \eqref{eq:flxle}. As shown in Figure \ref{fig:swiftuvot}, in the v and b bands, the results using the two approaches are almost identical. In the UV bands, especially uvw1 and uvw2, the light curve differences computed by the two approaches are substantial. This is because that the interstellar dust optical properties have more significant spectral dependence in the UV wavelengths. As shown in Figure \ref{fig:dustp}, the dust SSA have sharper variation with wavelength in UV. Thus, to accurately model light echoes in UV, the UV spectra of supernova are necessary.

The strong spectral dependence of UV dust optical properties does not impair the importance of obtaining SN~1987A early UV radiation information from the UV light echo. The strong UV breakout at the SN~1987A explosion early stage generates a UV light echo close to the impulse response of the dust cloud. Because SN~1987A V band light curve does not have an early peak, the differences between a UV and V band light echo light curve shapes contain the information of the duration and intensity of UV breakout. As shown in the inset plot of Figure \ref{fig:swiftuvot}, the early UV peaks at u and uvw1 bands are wider than at uvw2 and uvm2. This is reflected in the light echo light curves. Although the largest values of UV peaks are similar in the four UV bands, the light curve plateau magnitudes at u and uvw1 are about two magnitudes larger than at uvw2 and uvm2.

Further distinctive features can be measured if the echo can be spatially resolved. The echoes in the UV are in general much sharper than the echoes in the optical range and are located at the leading edges of the echoes.

\section{SUMMARY} \label{sec:summary}

In this study, we use an efficient MCRTM to simulate an SN~1987A light echo which was observed as AT2019xis in LMC, including light curves, light echo images and linear polarization. The theoretical simulations are compared with OGLE-IV OTDS light curve observations and are used to estimate the properties of the dust cloud that produced the AT2019xis light echo.

HST images show that the dust cloud in AT2019xis is a branch of a larger cloud and has a complicated shape. Because we do not know the exact shape of the dust cloud and for simplicity, we model the dust cloud as a homogeneous sphere or cube in simulations. The distance between the dust cloud and SN~1987A is estimated to be 186.3 pc using their angular distance and the delay time of the light echo. The single-scattering albedo and asymmetry factor data of six WD01 dust models are used in the simulations. The H-G phase function and WD01 asymmetry factor data are combined to obtain the dust cloud’s phase function or element (1,1) of the scattering phase matrix. The ratios of other scattering phase matrix elements to the phase function are the same as with the counterpart of the Rayleigh scattering phase matrix.

The light echo light curve magnitude and shape are sensitive to the dust cloud size, shape, optical thickness and dust grain optical properties. Because the dust cloud is absorptive, the light curve magnitude is not monotonic with an increasing optical thickness. With all other conditions unchanged, at any delay time, the light curve magnitude first increases with a larger optical thickness. If the optical thickness further increases, the light curve magnitude starts to decrease. The light echo light curve maximum arrives earlier if a dust cloud has a larger optical thickness. After the maximum, the light curve of an optically thick cloud drops faster than with an optically thin cloud. The linear polarization degree decreases with increased optical thickness, and its preferential polarization direction is always perpendicular to the global scattering plane. A larger-sized dust cloud scatters more light to the observer so the light curve magnitude increases monotonically with increased dust cloud size. The degree of linear polarization is not sensitive to the dust cloud size. The dust cloud shape mainly affects the light echo light curve shape around its global maximum. Around the maximum, a spherical dust cloud has a curved light curve, whereas a cube-shaped dust cloud has a flat light curve.

The time variation of light echo images of optically thick and thin clouds is very different. Because the dust cloud is absorptive, if the incident light is scattered too many times, most of the light energy is absorbed and cannot be observed in the light echo. Therefore, only light encountering a little attenuation in the dust cloud can be detected. For an optically thin cloud, the observer can see that the whole cloud is illuminated. For an optically thick cloud, only its optically thin edges are visible to the observer. An optically thick spherical cloud looks like a ring and an optically thick cube-shaped cloud only has its bottom edge visible. The light echo image is also determined by the SN~1987A light curves. The maximum of the SN~1987A light curve results in a local bright region in the light echo seen as a bright strip. The bright strip moves along with the light echo paraboloid. As a core-collapse supernova, SN~1987A has a strong and short UV-bright shock breakout just after the explosion, so its U-band light curve has an impulse-like feature before the maxima in other bands. Therefore, the U-band light echo has an impulse response-like feature, which distinguishes it from other bands’ light echoes.

The linear polarization pattern of a light echo is also strongly dependent on optical thickness. The light echo linear polarization pattern of an optically thin cloud spatially does not vary a lot, and its preferential polarization direction is perpendicular to the scattering plane. In contrast, the light echo linear polarization pattern of an optically thick cloud has a large spatial variation. For example, in a spherical dust cloud light echo, the linear polarization degree decreases in the regions closer to the light echo image center. Its preferential polarization direction is radial on the illuminated ring and is perpendicular to the local scattering plane. In an optically thin cloud, the observable light at a certain point of a light echo is an average of light experiencing various propagation paths. The spatial variation of linear polarization is thus insignificant. The averaged U Stokes parameter is zero because of symmetry. In an optically thick cloud, the observable light only focuses on the edge region and experiences similar paths. Thus, its linear polarization pattern keeps the local feature and has significant spatial variation.

We estimate the dust cloud size and optical thickness using the OGLE-IV OTDS AT2019xis I-band light curve observation and MCRTM simulations. The Levenberg-Marquardt method is utilized to minimize the cost function Equation \eqref{eq:costf}. The simulated light curves assuming various dust cloud optical properties and shapes all can roughly fit the observation well, even though the simulated light curves cannot capture some detailed light curve features. We also calculate the mass of the dust cloud using the estimated size, optical thickness and extinction coefficient. With an LMC gas-to-dust ratio of 300, the total mass of the cloud is about 7.8-9.3 \(M_\odot\). We use one of the best-fit dust cloud optical thickness and size value to calculate the light echo light curves in the Swift UVOT bands with the SN~1987A spectra. It is found that the photometry data of supernova are not enough to accurately model light echoes in UV bands, since interstellar dust optical properties vary significantly in UV wavelengths, so the UV spectra of supernova are necessary in modeling supernova UV light echoes.


\acknowledgments

This study was supported by the National Science Foundation (Grant No. AST-1817099). JD and PB were partially supported by NASA Astrophysics Data Analysis grant NNX17AF43G.  The computations were conducted at the Texas A\&M University Supercomputing Facility. This research has made use of the SVO Filter Profile Service (\url{http://svo2.cab.inta-csic.es/theory/fps/}) supported from the Spanish MINECO through grant AYA2017-84089. We thank Nicholas Suntzeff for helpful discussions. We also thank Steven R. Schroeder for helping checking the grammar.

\vspace{5mm}
\facilities{HST(STIS), OGLE-IV OTDS}
\software{astropy \citep{2013A&A...558A..33A}}



\clearpage
\appendix
\section{HST Image Data Information}\label{apdx:HSTdata}

\begin{deluxetable*}{lccr}[!ht]
\tablenum{A1}
\tablecaption{Pre-Echo Hubble Space Telescope Imaging\label{tab:hst}}
\tablewidth{0pt}
\tablehead{
\colhead{Dataset ID} & \colhead{Filter} & \colhead{Start Time (UTC)} &
\colhead{Exposure Time (s)}}
\startdata
ICD601BMQ & F475W & 2014-08-16 11:13:44 &  600 \phantom{aaaaaa} \\
ICD602DDQ & F475W & 2014-08-16 15:59:09 &  600 \phantom{aaaaaa} \\
ICD603GDQ & F475W & 2014-08-17 06:15:23 &  600 \phantom{aaaaaa} \\
ICD604I4Q & F475W & 2014-08-17 11:00:48 &  600 \phantom{aaaaaa} \\
ICD605LPQ & F475W & 2014-08-20 04:06:25 &  600 \phantom{aaaaaa} \\
ICD6010J0 & F606W & 2014-08-16 09:57:21 &  980 \phantom{aaaaaa} \\
ICD6020J0 & F606W & 2014-08-16 14:43:48 &  980 \phantom{aaaaaa} \\
ICD6030J0 & F606W & 2014-08-17 04:40:21 &  980 \phantom{aaaaaa} \\
ICD6040J0 & F606W & 2014-08-17 09:50:03 &  980 \phantom{aaaaaa} \\
ICD6050J0 & F606W & 2014-08-20 02:50:16 &  980 \phantom{aaaaaa} \\
ICD6010L0 & F814W & 2014-08-16 10:24:53 &  1230 \phantom{aaaaaa} \\
ICD6020L0 & F814W & 2014-08-16 15:11:20 &  1230 \phantom{aaaaaa} \\
ICD6030L0 & F814W & 2014-08-17 05:07:53 &  1230 \phantom{aaaaaa} \\
ICD6040L0 & F814W & 2014-08-17 10:17:35 &  1230 \phantom{aaaaaa} \\
ICD6050L0 & F814W & 2014-08-20 03:32:44 &  1230 \phantom{aaaaaa}
\enddata
\end{deluxetable*}

\section{Monte Carlo Radiative Transfer Model}\label{apdx:MCRTM}

The MCRTM is based on a forward photon tracing technique. Here ‘a photon’ represents a very small portion of light power emitted by the source (e.g., SN~1987A). A random number determines the photon scattering location, and its scattering direction is estimated according to the scattering phase function. The individual photon’s total path length and Stokes vector are monitored.

The position of a photon is expressed in Cartesian coordinates as
\begin{equation} \label{eq:pos}
\vec{r}=r\begin{pmatrix} \sin{{\theta}_r}\cos{{\phi}_r}, & \sin{{\theta}_r}\sin{{\phi}_r}, & \cos{{\theta}_r} \end{pmatrix},
\end{equation}

\noindent
where ${\theta}_r$ and ${\phi}_r$ are zenith and azimuth angles of the position vector in laboratory coordinate respectively, and $r$ is the distance between the photon and coordinate origin. Laboratory coordinates refer to a fixed $xyz$ Cartesian coordinate system. Here, the origin $O$ is at the light source, the $z$-axis is along the line of sight to the observer, and the $xy$-axes are perpendicular to $z$ at the light source. Coordinates of all photon propagation and scattering events are expressed in this fixed system without transforming axes as each photon changes direction. The photon’s propagation direction is expressed as
\begin{equation} \label{eq:pdir}
\hat{e}=\begin{pmatrix} \sin{{\theta}_e}\cos{{\phi}_e}, & \sin{{\theta}_e}\sin{{\phi}_e}, & \cos{{\theta}_e} \end{pmatrix},
\end{equation}

\noindent
where ${\theta}_e$ and ${\phi}_e$ are zenith and azimuth angles of the directional vector in laboratory coordinates.

In the MCRTM, a scattering event occurs at the photon’s position $\vec{r}$. From current scattering position $\vec{r}_s$ to the next scattering position $\vec{r}_{s+1}$, the pathlength is $d_s$ and satisfies
\begin{equation} \label{eq:dses}
d_s\hat{e}_s=\vec{r}_{s+1}-\vec{r}_s,
\end{equation}

\noindent
where the subscript $'s'$ denotes the scattering order. The pathlength is obtained as
\begin{equation} \label{eq:taus}
{\tau}_s=\int_{0}^{d_s}C^s_\text{ext}(r')\,dr'=-\ln{\left[1-{\xi}_1\left(1-e^{-{\tau}_t}\right)\right]},
\end{equation}

\noindent
where ${\tau}_s$ is the optical thickness from $\vec{r}_s$ to $\vec{r}_{s+1}$, $C^s_\text{ext}$ is the extinction coefficient along the path from $\vec{r}_s$ to $\vec{r}_{s+1}$, ${\tau}_t$ is the optical thickness along $\hat{e}_s$ from $\vec{r}_s$ to the boundary of the scattering medium, and ${\xi}_1$ is a random number between 0 and 1. Equation \eqref{eq:taus} guarantees $\vec{r}_{s+1}$ is always in the scattering medium and imposes a weight $w_s=1-e^{-{\tau}_t}$ on this photon’s contribution.

At $\vec{r}_{s+1}$, the new propagation direction of the photon is determined by $\hat{e}_s$, scattering angle ${\Theta}_{s+1}$ and azimuth angle ${\Phi}_{s+1}$. ${\Theta}_{s+1}$ and ${\Phi}_{s+1}$ are angles in a local coordinate where $\hat{e}_s$ is in local $z$ direction along the preceding propagation direction. ${\Theta}_{s+1}$ and ${\Phi}_{s+1}$ are computed as
\begin{equation} \label{eq:Theta}
\frac{\int_0^{{\Theta}_{s+1}}F_{11}(\vec{r}_{s+1},{\Theta}')\,\sin{{\Theta}'}\,d{\Theta}'}{\int_0^{\pi}F_{11}(\vec{r}_{s+1},{\Theta}')\,\sin{{\Theta}'}\,d{\Theta}'}={\xi}_2,
\end{equation}
\begin{equation} \label{eq:Phi}
{\Phi}_{s+1}=2\pi{{\xi}_3},
\end{equation}

\noindent
where $F_{11}$ is the scattering phase function or the (1,1) element of the scattering phase matrix, and ${\xi}_2$ and ${\xi}_3$ are two independent random numbers with probability of a uniform distribution between 0 and 1. The zenith and azimuth angles of $\hat{e}_{s+1}$ in laboratory coordinates are
\begin{equation} \label{eq:theta}
{\theta}^{s+1}_e=\arccos{\left(\cos{{\Theta}_{s+1}}\cos{{\theta}^{s}_e}
+\sin{{\Theta}_{s+1}}\sin{{\theta}^{s}_e}\cos{{\Phi}_{s+1}}\right)},
\end{equation}
and
\begin{equation} \label{eq:phi}
{\phi}^s_e={\Theta}_{s+1}+\arccos{\left(\frac{\cos{{\Theta}_{s+1}}-\cos{{\theta}^{s+1}_e}\cos{{\theta}^{s}_e}}{\sin{{\theta}^{s+1}_e}\sin{{\theta}^{s}_e}}\right)}.
\end{equation}

The photon’s Stokes vector at $\vec{r}_{s+1}$ is updated as
\begin{multline} \label{eq:stks}
\begin{pmatrix} I\\Q\\U\\V\end{pmatrix}_{s+1}=w_s{\varpi}\left(\vec{r}_{s+1}\right){\times}\\
\mathbf{L}\left(\pi-{\sigma}_2\right)\mathbf{F}\left(\vec{r}_{s+1},{\Theta}_{s+1}\right)\mathbf{L}\left(-{\sigma}_1\right)\begin{pmatrix} I\\Q\\U\\V\end{pmatrix}_{s},
\end{multline}

\noindent
where $I$, $Q$, $U$ and $V$ are Stokes vector elements, $\mathbf{F}$ is the scattering phase matrix, $\varpi$ is the single-scattering albedo (SSA) and $\mathbf{L}$ is the rotation matrix in the form

\begin{equation} \label{eq:rotm}
\begin{pmatrix}
1 & 0 & 0 & 0 \\
0 & \cos{2{\sigma}} & \sin{2{\sigma}} & 0 \\
0 & -\sin{2{\sigma}} & \cos{2{\sigma}} & 0 \\
0 & 0 & 0 & 1
\end{pmatrix}.
\end{equation}

$\mathbf{L}(-{\sigma}_1)$ rotates the polarization reference plane from the incident meridional plane (formed by $\hat{e}_{s}$ and laboratory $z$ direction) to the scattering plane (formed by $\hat{e}_{s}$ and $\hat{e}_{s+1}$). $\mathbf{L}(\pi-{\sigma}_2)$ rotates the polarization reference plane from the scattering plane to the scattering meridional plane (formed by $\hat{e}_{s+1}$ and laboratory $z$ direction). The angles ${\sigma}_1$ and ${\sigma}_2$ are thus derived as
\begin{subequations} \label{eq:sigma}
\begin{equation}
{\sigma}_1=\arccos{\left(\frac{\cos{{\theta}^{s+1}_e}-\cos{{\Theta}_{s+1}}\cos{{\theta}^{s}_e}}{\sin{{\Theta}_{s+1}}\sin{{\theta}^{s}_e}}\right)},
\end{equation}
\begin{equation}
{\sigma}_2=\arccos{\left(\frac{\cos{{\theta}^{s}_e}-\cos{{\Theta}_{s+1}}\cos{{\theta}^{s+1}_e}}{\sin{{\Theta}_{s+1}}\sin{{\theta}^{s+1}_e}}\right)}.
\end{equation}
\end{subequations}

In the traditional forward photon tracing technique, only the photons exiting the scattering medium and propagating to the observational directions are counted as contributions to the simulated observed signal. Traditional forward photon tracing is inefficient because it usually needs more than ten million photons to achieve a convergent result and sometimes still contains noticeable random noise. To improve efficiency (reducing the required photons to about one-tenth), in this MCRTM, at every scattering event, we consider the contribution of a photon to the simulated observation signal.

At any scattering position $\vec{r}_{s+1}$, the photon’s contribution to the simulated observation signal Stokes vector is computed as
\begin{multline} \label{eq:stko}
\begin{pmatrix} I\\Q\\U\\V\end{pmatrix}_{o}=\frac{w_s}{4\pi}\exp{\left(-{\tau}'_t\right)}{\varpi}\left(\vec{r}_{s+1}\right){\times}\\
\mathbf{L}\left(\pi-{\sigma}'_2\right)\mathbf{F}\left(\vec{r}_{s+1},{\Theta}_{o}\right)\mathbf{L}\left(-{\sigma}'_1\right)\begin{pmatrix} I\\Q\\U\\V\end{pmatrix}_{s},
\end{multline}

\noindent
where subscript $'o'$ denotes observation, the scattering angle ${\Theta}_{o}$ is between $\hat{e}_{s}$ and observer direction $\hat{e}_{o}$ (i.e., $\cos{{\Theta}_{o}}=\hat{e}_{s}\cdot\hat{e}_{o}$), ${\sigma}'_1$ and ${\sigma}'_2$ are determined in the same way as in Equation \eqref{eq:sigma}, and ${\tau}'_t$ is the optical thickness from the scattering location to the observer. In one scattering event, we can specify different observer directions and obtain the photon’s contribution to multiple directions. With the same number of simulated photons, the Monte Carlo photon tracing converges much faster than the traditional approach, if Equation \eqref{eq:stko} is used to count photon contributions.

At a scattering position $\vec{r}_{s}$, the photon’s total path length is
\begin{equation} \label{eq:dtot}
d^{tot}_s=\sum_{i=1}^sd_i.
\end{equation}

The delay time relative to the time when the source light directly propagates to the observer is derived according to a geometric relation \citep{Patat2005}:
\begin{equation} \label{eq:tds}
t^d_s=\frac{d_{tot}-\hat{e}_{o}\cdot\vec{r}_{s}-l_{\text{SN}}}{c},
\end{equation}

\noindent
where $c$ is the speed of light, and $l_{\text{SN}}$ is the distance between the observer and source. In the MCRTM, delay time bins are predefined with spacing ${\Delta}t$. If $t^d_s$ is within a predefined time bin, the photon’s contribution is counted toward the light echo at the center of this time bin.

The observed Stokes vector computed by Equation \eqref{eq:stko} must be normalized by the total emitted flux within the solid angle of the dust cloud viewed from SN~1987A. The normalization factor $F_0$ is defined as
\begin{equation} \label{eq:f0}
F_0=\frac{n_\text{ph}F_\text{SN}}{\Delta{\Omega}_\text{ds}{\Delta}t},
\end{equation}

\noindent
where $n_\text{ph}$ is the total number of emitted photons, $F_\text{SN}$ is the flux of the supernova and $\Delta\Omega_\text{ds}$ is the solid angle of the dust cloud viewing from the supernova. Since the scattered flux is assumed to be linearly proportional to the source flux, by setting $F_\text{SN}$ to be one, we can directly multiply the radiative transfer result by the incident flux to obtain the scattered flux for a specific source. In the model, the supernova is taken to be a point light source. If the dust cloud is a sphere, $\Delta\Omega_\text{ds}$ is computed as
\begin{equation} \label{eq:dOmg}
\Delta\Omega_\text{ds}=2\pi\left(1-\frac{\sqrt{l^2_\text{ds}-R^2_\text{d}}}{l_\text{ds}}\right),
\end{equation}

\noindent
where $R_\text{d}$ is the dust cloud sphere radius. To guarantee that all the emitted photons propagate to the dust cloud, the photon’s initial propagation direction is confined within $\Delta\Omega_\text{ds}$.

\bibliography{references}{}
\bibliographystyle{aasjournal}



\end{document}